\documentclass[12pt,preprint]{aastex}

\slugcomment{Not to appear in Nonlearned J., 45.}

\shorttitle{Spectral Line Survey toward M51}
\shortauthors{Watanabe et al.}

\begin{document}


\title{Spectral Line Survey toward Spiral Arm of M51 \\in the 3~mm and 2~mm Bands}


\author{Yoshimasa~Watanabe\altaffilmark{1,2}, Nami~Sakai\altaffilmark{1}, Kazuo~Sorai\altaffilmark{3},}

\and

\author{Satoshi Yamamoto\altaffilmark{1}}
\affil{\altaffilmark{1}Department of Physics, The University of Tokyo, 7-3-1, Hongo, Bunkyo-ku, Tokyo, 113-0033, Japan}
\affil{\altaffilmark{3}Department of Physics / Department of Cosmoscience, Hokkaido University, Kita 10, Nishi 8, Kita-ku, Sapporo, Hokkaido, 060-0810, Japan}





\begin{abstract}
We have conducted a spectral line survey in the 3~mm and 2~mm bands toward two positions in a spiral arm of M51 (NGC~5194) with the IRAM 30~m telescope.  In this survey, we have identified 13 molecular species, including CN, CCH, N$_2$H$^+$, HNCO, and CH$_3$OH.  Furthermore, 6 isotopologues of the major species have been detected. On the other hand,  SiO, HC$_3$N, CH$_3$CN, and the deuterated species such as DCN and DCO$^+$ are not detected.  The deuterium fractionation ratios are evaluated to be less than 0.8~\% and 1.2~\% for DCN/HCN and DCO$^+$/HCO$^+$, respectively.  By comparing the results of the two positions with different star formation activities, we have found that the observed chemical compositions do not strongly depend on star formation activities.  They seem to reflect a chemical composition averaged over the 1-kpc scale region including many giant molecular clouds.  Among the detected molecules CN, CCH, and CH$_3$OH are found to be abundant.  High abundances of CN, and CCH are consistent with the above picture of a wide spread distribution of molecules, because they can be produced by photodissociation.  On the other hand, it seems likely that CH$_3$OH is liberated into the gas phase by shocks associated with large scale phenomena such as cloud-cloud collisions and/or by non-thermal desorption processes such as photoevaporation due to cosmic-ray induced UV photons.  The present result demonstrates a characteristic chemical composition of a giant molecular cloud complex in the spiral arm, which can be used as a standard reference for studying chemistry in AGNs and starbursts.
\end{abstract}





\section{Introduction}

As inferred from astrochemical studies for Galactic sources, chemical compositions often tell us much more about physical conditions of target sources than conventional physical parameters such as mass, density, and temperature.  For example, dynamical evolution of starless cores has been studied on the basis of chemical indicators such as the CCS/NH$_3$ ratio and the deuterium fractionation ratio \citep[\textit{e.g.}][]{suzuki1992, Caselli2003, Aikawa2013}. This direction will be more and more important in extragalactic studies in the near future.  However, efforts interpreting chemical compositions in terms of physical conditions are still in their infancy for external galaxies.  Their chemical compositions have often been discussed on the basis of astrochemical concepts established in nearby molecular clouds in our Galaxy without careful considerations on a large difference of a resolvable scale between Galactic molecular clouds and external galaxies.  For external galaxies, we usually observe the `average' chemical compositions of a number of giant molecular clouds (GMCs).  Although this situation is being improved by high angular resolution observations with interferometers, there still remains a large gap between extragalactic chemistry and molecular-cloud chemistry.  Therefore, it is of fundamental importance to understand the chemical composition averaged over GMCs and its meanings.  This cannot be done with models, but has to be done with observations.

Chemical compositions of external galaxies are mainly studied in the nuclear regions.  They are known to vary under extreme environments such as the starbursts and the AGNs.  For example, it has been suggested that the abundance ratio of HCN and HCO$^{+}$ depends on relative contribution of photodissociation regions (PDRs) and X-ray dominated regions (XDRs) \citep[\textit{e.g.}][]{Kohno2001,Lepp1996,Maloney1996,Meijerink2005,Meijerink2007}.  Unbiased spectral line surveys in a wide frequency range became realistic for external galaxies, thanks to recent technical developments of sensitive receivers and wide-band correlators.  Many spectral line surveys have been conducted toward the nuclear regions of nearby galaxies with single dish telescopes \citep[\textit{e.g.}][]{martin2006, aladro2011b, aladro2013, costagliola2011, nakajima2011} and even with an interferometer \citep[\textit{e.g.}][]{martin2011}.  In order to extract chemical features associated with the extreme environments of the nuclear regions, it is indispensable to know the chemical composition averaged over the similar size including a number of GMCs in a spiral arm as a reference.  However, only little efforts have been done for such a fundamental task.  For instance, \citet{meier05, meier2012} studied distributions of CCH, C$^{34}$S, N$_2$H$^+$, CH$_3$OH, HNCO, HNC, HC$_3$N, and SO in IC~342 and Maffei~2 with the OVRO and BIMA mm-interferometers.  They observed a part of the spiral arm region as well as the nuclear region, and revealed that the distributions are different from molecule to molecule.  Despite such an effort, we still need to explore the meaning of the chemical composition averaged over the 1-kpc scale region involving many GMCs in a spiral arm, which would have a characteristic chemical composition as a whole.  If we can understand its meanings in relation to physical conditions and internal structures, it could be used as a standard reference for chemical diagnostics of external galaxies.  As the first step toward this goal, we have conducted an unbiased spectral line survey toward a spiral arm in M51 to characterize the averaged chemical composition of GMCs there.

M51 (NGC~5194) is a nearby grand-design spiral galaxy (SAbc) with a distance of 8.4~Mpc \citep{feldmeier1997,vink12}.  Molecular gas distribution and kinematics of this galaxy have extensively been studied with CO emission \citep{nakai1994,aalto1999,helfer2003,schuster2007,koda09,egusa11,schinnere13}.  In addition, HCN \citep{nguyen1992,kohno1996,sorai2002}, HCO$^{+}$ \citep{nguyen1992}, HNC \citep{huettemeister1995}, CS \citep{sage1990}, N$_2$H$^+$ \citep{sage1995}, and NH$_3$ \citep{takano2013} have been detected toward the nuclear region.  In spiral arm regions, \citet{schinnere10} observed the $^{12}$CO, $^{13}$CO, C$^{18}$O, HCO$^+$, and HCN lines.  They have estimated kinetic temperature and H$_2$ density by LVG analysis, and suggested that star formation activities do not have a strong impact on physical conditions of molecular gas probed at their observational resolution of 120~pc.  However, the observation is limited to a few representative molecular species.  In this study, we have investigated the chemical composition of GMCs averaged over the 1-kpc scale region toward two positions in a spiral arm with different star formation activities by the spectral line survey.

\section{Observations}
\begin{figure}
\epsscale{1.00}
\plotone{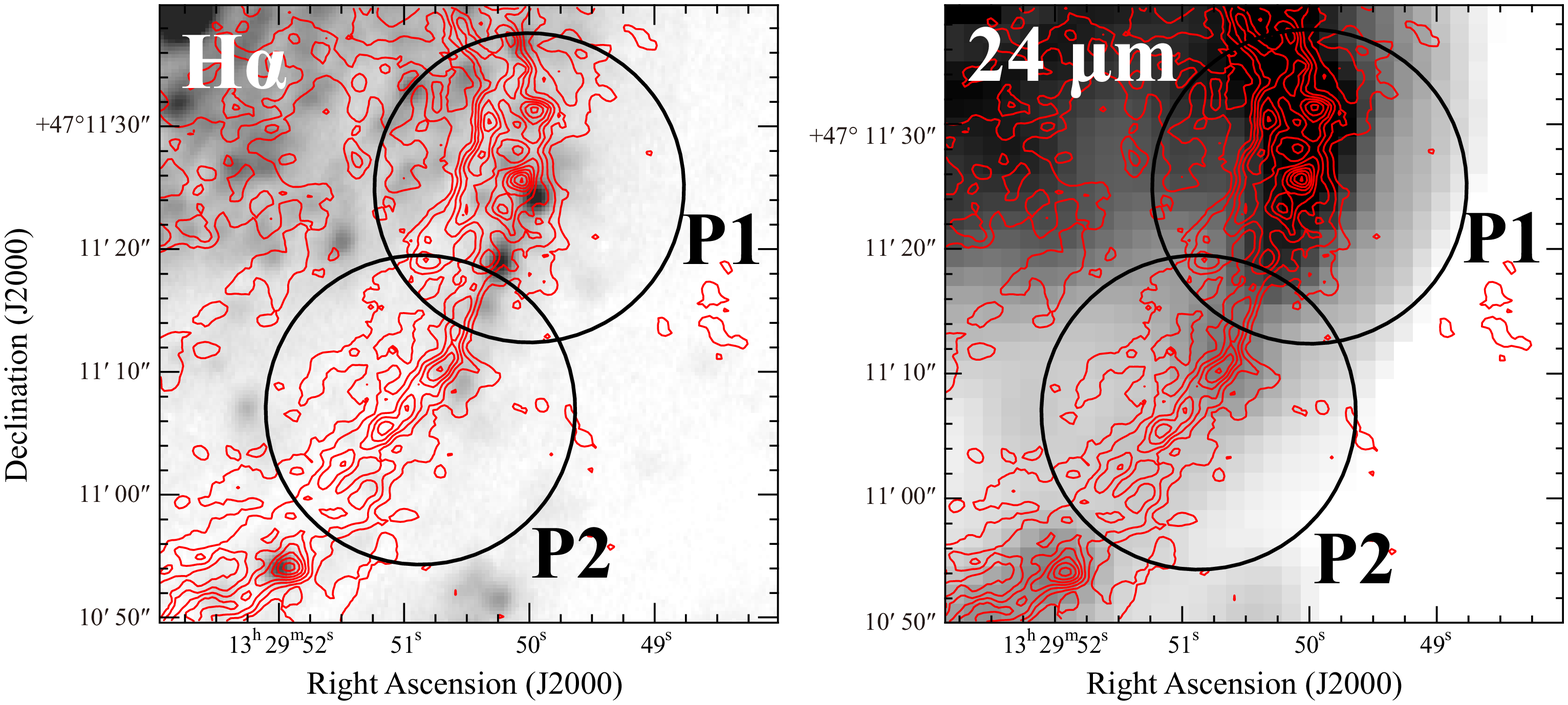}
\caption{Observed positions (black circles) on the H$\alpha$ image (a) and the 24~$\mu$m image (b) by SINGS \citep{kennicutt2003}.  The red contours are the $^{12}$CO intensities by PAWS \citep{schinnere13}.  The contour levels are from 40~K~km~s$^{-1}$ to 440~K~km~s$^{-1}$ with an interval of 50~K~km~s$^{-1}$.}
\label{fig0}
\end{figure}

Observations were performed with the IRAM~30~m telescope at Pico Veleta in 23--29, December, 2011 and 10--13, August, 2012.  The observed positions were ($\alpha_{\rm J2000}$, $\delta_{\rm J2000}$) = (13$^{\rm h}$29$^{\rm m}$50$^{\rm s}$.0, +47$^{\circ}$11$^{'}$25$^{''}$.0) (P1) and ($\alpha_{\rm J2000}$, $\delta_{\rm J2000}$) = (13$^{\rm h}$29$^{\rm m}$50$^{\rm s}$.9, +47$^{\circ}$11$^{'}$06$^{''}$.9) (P2) (Figure \ref{fig0}).  These are the $^{13}$CO($J=1-0$) peak positions of the spiral arm in M51 \citep{schinnere10}.  P1 contains many of the Balmer~$\alpha$ and Paschen~$\alpha$ spots and shows strong 24 micron emission, while these emissions are generally weaker in P2 than in P1 \citep{schinnere10}.  Therefore, star formation activity is higher in P1 than in P2.  The beam size is 30--20$^{''}$ and 19--17$^{''}$ at the 3~mm and 2~mm band, respectively, which corresponds to a linear scale of 1.2--0.9~kpc and 0.8--0.7~kpc, respectively.  The observed frequency range is from 83 to 103.5, from 108.5 to 116~GHz, and from 130 to 148~GHz for P1, and from 85 to 116~GHz and from 139.3 to 142.3~GHz for P2.  Since the frequency range from 100 to 109~GHz does not contain strong spectral lines according to spectral line surveys toward the Galactic sources, we put a lower priority on observations of this frequency range within the limited observation time.  Two EMIR (Eight MIxer Receivers) bands, E090 and E150, were observed simultaneously with the dual polarization mode.  The EMIR is the sideband separating receiver, which outputs the USB and LSB signals separately.  The image rejection ratios are assured to be better than 13~dB and 10~dB for E090 and E150, respectively, according to the status report of the 30~m telescope\footnote{\url{http://www.iram.es/IRAMES/mainWiki/EmirforAstronomers}}.  The system temperatures ranged from 75-310~K for E090 and from 100-210~K for E150.  The detailed frequency settings, integration time, and system noise temperatures are summarized in Table \ref{tab7}.  The backends were eight FTS (Fourier Transform Spectrometers) autocorrelators whose bandwidth and channel width are 4050~MHz and 195~kHz, respectively.  The telescope pointing was checked every hour by observing continuum sources near the target position, and was found to be better than $\pm 5^{''}$.  The wobbler switching mode was employed with beam throw of $\pm120^{''}$ and switching frequency of 0.5~Hz.  The wobbler throw is toward the azimuth direction, and hence, the absolute off-position depends on the hour angle.  Although the off-position is still within the M51 system, the emission of the off-position is smeared out for the long integrations in our line survey observation.  In fact, we compared the CO intensity with that of obtained by PAWS \citep{schinnere13}, and found that the off-position contribution is less than 10~\%.  Hence, the effect would be almost negligible for the other molecular lines with higher critical densities.  The intensity scale was calibrated to the antenna temperature ($T_{\rm A}^{*}$) scale by using cold and hot loads.  $T_{\rm A}^{*}$ was converted to the main beam temperature $T_{\rm mb}$ by multiplying $F_{\rm eff}/B_{\rm eff}$.  Here, $F_{\rm eff}$ is forward efficiency, which is 95~\% and 93~\% for the 3~mm and 2~mm bands, respectively, and $B_{\rm eff}$ is main beam efficiency, which is 81~\% and 74~\% for the 3~mm and 2~mm bands, respectively.  Thanks to the wobbler switching, the spectrum baseline was stable.  Spectral baselines were subtracted for each correlator band (4050~MHz) and for each scan by fitting a 5th - 7th order polynomial to the line-free part before the averaging of all the scans.  Then, the final spectrum for each correlator band was obtained by integrating all the scans.  The composite spectrum was prepared by gathering the spectra thus obtained.  

\section{Results}
\subsection{Detected Molecules}
\begin{figure}
\epsscale{1.00}
\plotone{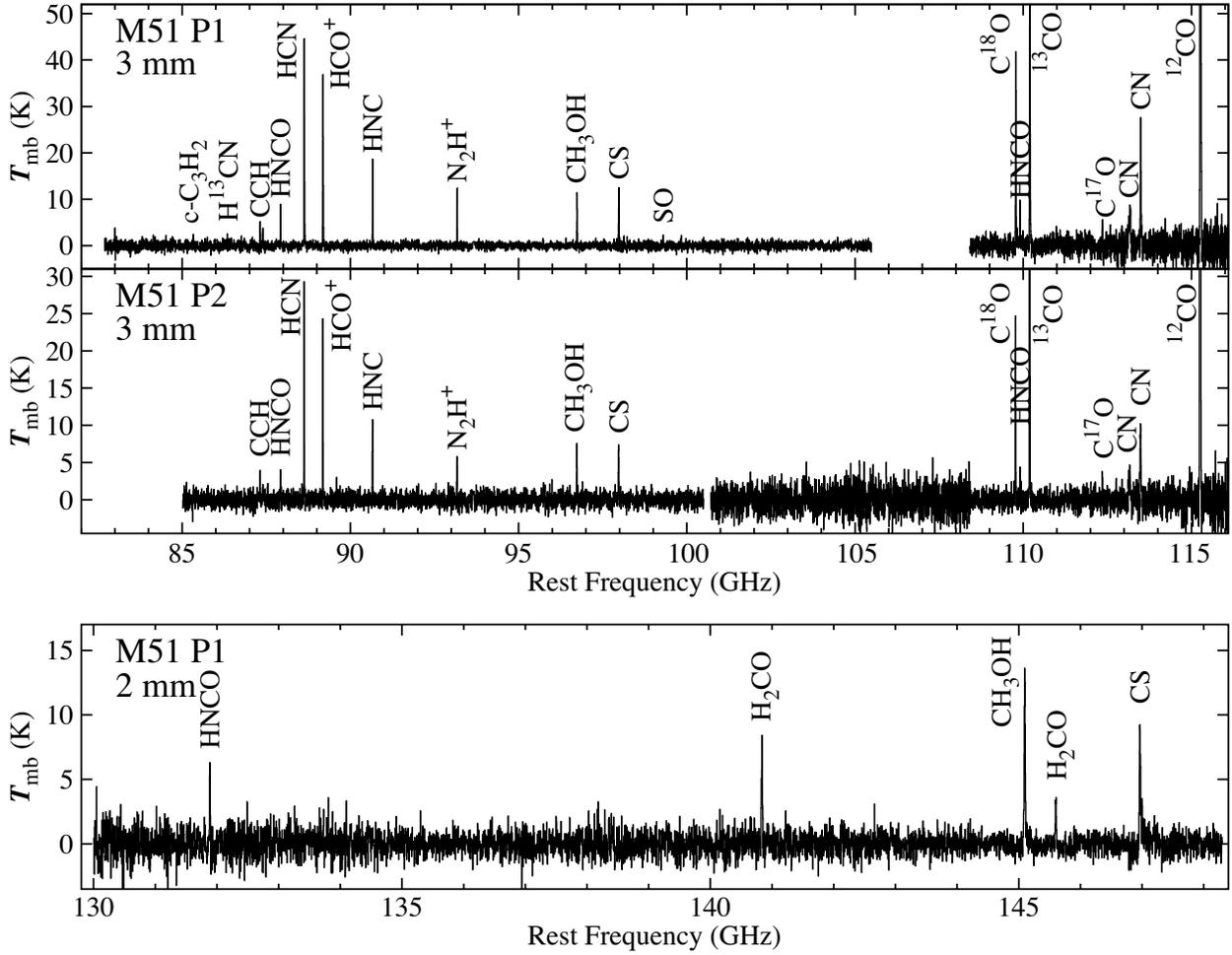}
\caption{Compressed spectra of M51~P1 (Top) and M51~P2 (Middle) in the 3~mm band, and that of P1 in the 2~mm band (Bottom).  Frequency resolution is 4~MHz. }
\label{fig1}
\end{figure}

\begin{figure}
\epsscale{0.97}
\plotone{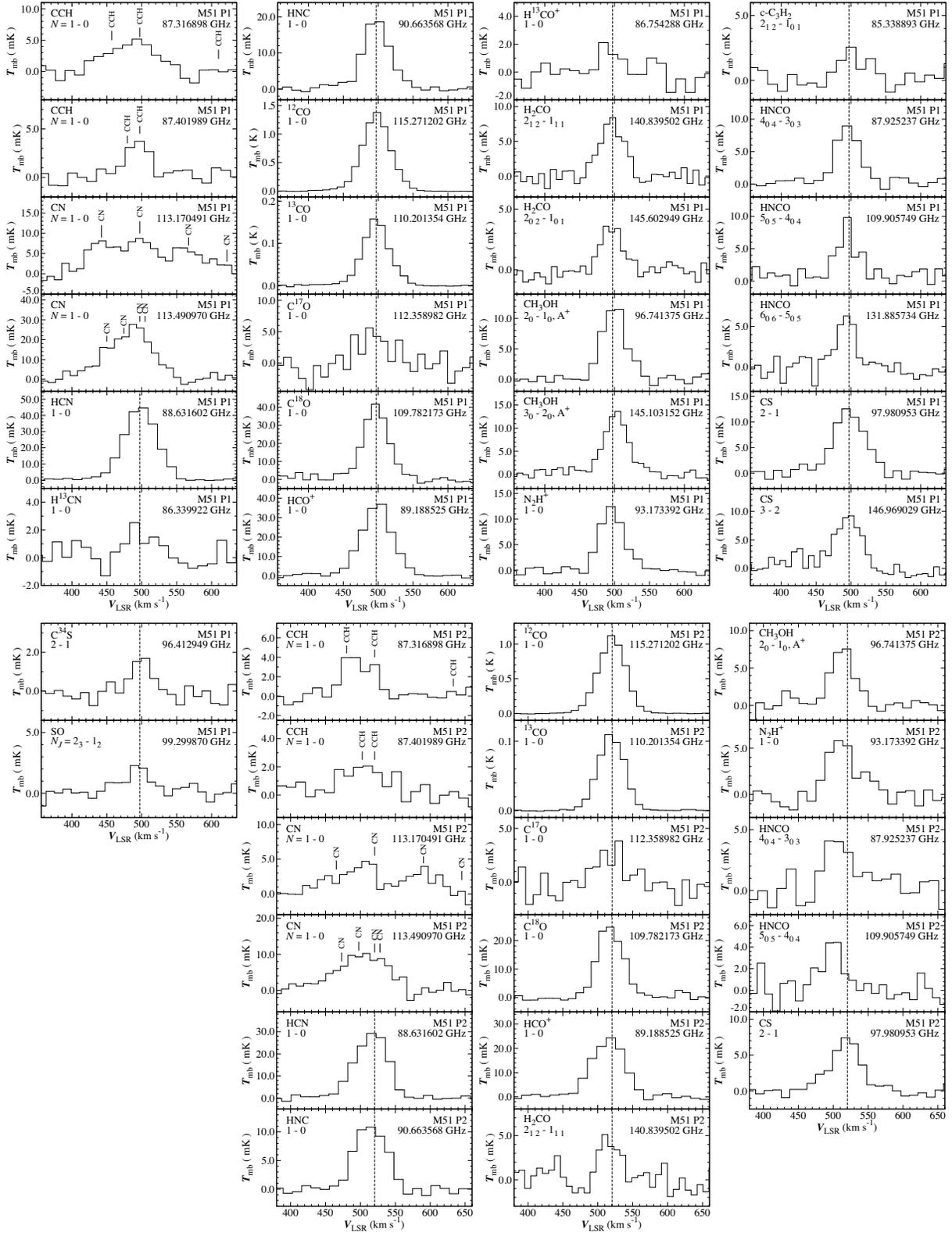}
\caption{Profiles of individual molecular lines observed in M51~P1 and P2.  Frequency resolution is 4~MHz, which corresponds to 14, 12, and 8~km~s$^{-1}$ for 85, 100, and 145~GHz, respectively.  Vertical lines represent the $V_{\rm LSR}$ of the $^{13}$CO line as reference.  They are not given for lines with hyperfine components or internal rotation components.  }
\label{fig3}
\end{figure}

Figure \ref{fig1} shows the composite spectra from 85 to 116~GHz toward P1 and P2 and that from 130 to 148~GHz in P1.  Their expanded spectra are shown in Appendix A.  Typical r.m.s. noise levels are 2--7~mK and 2--3~mK in the 3~mm and 2~mm bands, respectively for the spectral resolution of 4~MHz.  In total, 21~lines and 5~lines were detected in the 3~mm and 2~mm bands, respectively, toward P1 (Figure \ref{fig3} and Table \ref{tab1}), from which 13 molecular species and 6 isotopologues were identified with the aid of the spectral line database the Cologne Database for Molecular Spectroscopy (CDMS) managed by University of Cologne \citep{muller01,muller05} and Submillimeter, Millimeter, and Microwave Spectral Line Catalog provided by Jet Propulsion Laboratory \citep{{pickett98}}.  The line parameters were obtained by Gaussian fitting to the appropriate frequency range around the line which is much wider than the line width (Table \ref{tab1}).  A criterion of the detection is that the line is observed at the expected velocity ($497\pm20$ km~s$^{-1}$ and $520 \pm 20$ km~s$^{-1}$ for P1 and P2, respectively) with a line width of 20~km~s$^{-1}$ or broader and a significance of 3$\sigma$ or higher.  The result is shown in Table \ref{tab1}.  In addition to $^{12}$CO, HCN, HCO$^{+}$, HNC, and CS, which are usually found in the nuclear regions of many external galaxies, N$_{2}$H$^{+}$, H$_{2}$CO, CH$_{3}$OH, HNCO, CCH, CN, c-C$_{3}$H$_{2}$, and SO were detected toward P1.  As for isotopologues, H$^{13}$CN, H$^{13}$CO$^{+}$, and C$^{34}$S were identified, as well as $^{13}$CO, C$^{18}$O, and C$^{17}$O.  Toward P2, 16 and 1 lines were detected in the 3~mm and 2~mm bands, respectively (Figure \ref{fig3} and Table \ref{tab2}), from which 11 molecular species and 3 CO isotopologues have been identified.  Although the observation sensitivities are almost the same between P1 and P2, the number of detected lines is smaller in P2 than that in P1.  This is because the line intensities are about 1.5 times weaker in P2 than P1.

On the other hand, molecules often found in starburst galaxies and AGNs such as HC$_3$N, CH$_3$CN, CH$_{3}$CCH, OCS and SiO were not detected in the spiral arm of M51.  Although marginal features of the HC$_3$N line were recognized at the expected frequencies (90.979023~GHz and 100.076392~GHz for the $J=10-9$ and $J=11-10$ lines) in P1, they are not statistically significant ($<3\sigma$).  Moreover, deuterated species such as DCN, DCO$^{+}$, and HDCO were not detected.  Table \ref{tab1} also lists lines of representative non-detected species.  The upper limits to the peak intensities are 3 times the r.m.s. noise at the spectral resolution of 4~MHz, whereas the upper limits to the integrated intensities are evaluated from the upper limits to the peak intensities by assuming a velocity width of 40~km~s$^{-1}$.  In general, non-detection of the lines does not always mean low column densities, because the line may not be excited to give observable emission due to insufficient H$_2$ density and/or low kinetic temperature.  However, most of these non-detections do not originate from such an excitation problem, but are likely due to low abundances of these molecules.  Indeed, the HNCO line ($6_{0\,6} - 5_{0\,5}$) whose upper state energy is as high as 22.2~K was detected.  Upper state energies of transitions of the above non-detected molecules are lower than those of the detected lines (\textit{e.g.} HNCO), except for the HC$_3$N($J=10-9$) line ($E_{\rm u}=24.0$~K).  It should be noted that non-detections are based on the averaged over the telescope beam.  Hence, the results do not rule out the possibility that they exist in very localized parts within the beam.

\subsection{Excitation Temperatures and Molecular Abundances}
In this observation, multiple transition lines were detected for HNCO, CH$_3$OH, and CS.  Assuming an optically thin condition and local thermodynamic equilibrium (LTE), we determined the rotation temperatures and the beam-averaged column densities of these three molecules by using the least-square method with the following formula:
\begin{equation}
W_{\nu} = \frac{8 \pi^3 S\mu_0^2 \nu N}{3 k U(T_{\rm rot})} \left\{ 1 - \frac{\exp(h \nu/kT_{\rm rot})-1}{\exp(h \nu/kT_{\rm bg})-1} \right\}\exp\left(-\frac{E_{\rm u}}{k T_{\rm rot}}\right),
\end{equation}
where $W_{\nu}$, $S$, $\mu_0$, $\nu$, $N$, $k$, $U(T_{\rm rot})$, $T_{\rm rot}$, $h$, $T_{\rm bg}$, and $E_{\rm u}$ are integrated intensity, line strength, dipole moment, transition frequency, total column density, the Boltzmann constant, partition function, rotation temperature, the Planck constant, the cosmic microwave background temperature, and upper state energy, respectively.  Since we are looking at smaller regions than the telescope beam, we need to correct for the effect of beam dilution.  This correction is particularly important for using the 2~mm and 3~mm data simultaneously, because the beam dilution effect is very different.  The correction was made by dividing the observed integrated intensity by the beam dilution factor $(\theta_{\rm source}/\theta_{\rm beam})^2$, where $\theta_{\rm source}$ and $\theta_{\rm beam}$ is the source size and the $FWHM$ beam width of the telescope.  The $FWHM$ beam width can be calculated as $2.41 \times 10^3 / \nu ({\rm GHz})$ in arcsecond for the 30~m telescope.  Since, the size of the emitting region is uncertain and the emitting region may be different from molecule to molecule, we assumed three source sizes of 5$^{''}$, 10$^{''}$, and larger than the telescope beams.  The rotation temperature and the column density of CS, CH$_3$OH, and HNCO were estimated toward P1 as shown in Table \ref{tab3}.  In the case of CH$_3$OH, we need a special treatment, because the CH$_3$OH lines consist of unresolved multiple transitions (Table \ref{tab1}).  By assuming that the column density of the A-state $N$ is the same as that of the E-state, the intensity of CH$_3$OH can be expressed as: 
\begin{equation}
W_{\nu} = \sum_i\frac{8 \pi^3 S_i \mu_0^2 \nu_i N}{3 k U(T_{\rm rot},i)} \left\{ 1 - \frac{\exp(h \nu_i /kT_{\rm rot})-1}{\exp(h \nu_i /kT_{\rm bg})-1} \right\}\exp\left(-\frac{E_{{\rm u}i}}{k T_{\rm rot}}\right),
\end{equation}
Here, $i$ denotes an index of each transition, and $U(T_{\rm ex}, i)$ stands for the partition function of the A or E-state depending on the state that the $i$th line belongs to.  For P2, the rotation temperature was not determined accurately because of poor S/N ratios of the lines.  The rotation temperatures obtained toward P1 are similar to those found in cold dark clouds.  From this result, the detected molecules would reside in a cold ($\sim10$~K) and widespread molecular gas, although a part of the molecular emissions may also come from hot molecular gas affected by the feedback from the star formation activities.  Indeed, the spectral intensity patterns obtained in this survey are much different from those found in hot cores in high-mass star forming regions such as Orion~KL \citep{tercero2010}.  The observed spectra cannot be reproduced by an assembly of molecular gas in hot cores.  This is a natural consequence of the single dish observation which samples many GMCs within the telescope beam.  

Column densities of other molecules, for which only one line was observed, were estimated under the assumption of the LTE with a rotation temperature ($T_{\rm rot}$) of 5~K and 10~K by using eq. (1) (Table \ref{tab4}).  In addition to the identified molecules, upper limits to column densities were derived for 12 non-detected molecules (Table \ref{tab4}).  For H$_2$CO in P1, the column density was estimated by summing the column densities of ortho-H$_2$CO and para-H$_2$CO, which were derived from transitions of $2_{1\,2}-1_{1\,1}$ and $2_{0\,2}-1_{0\,1}$, respectively.  For H$_2$CO in P2, the ortho-to para ratio is assumed to be 3 (the statistical value).  The ortho-to-para ratios are also assumed to be 3 for c-C$_3$H$_2$, and H$_2$CS.  The upper limits for CH$_3$CCH and CH$_3$CN were evaluated by assuming equal column densities for the A and E states.  Fractional abundances were calculated by using the column density of molecular hydrogen derived from the column density of C$^{18}$O, as summarized in Table \ref{tab4}.  Here, the [H$_2$]/[C$^{18}$O] ratio of $2.9 \times 10^6$ \citep{meier05} is employed.  For simplicity, we assumed that all the molecular species have the same size of the emitting regions.  Although the column densities are strongly dependent on the assumed source sizes, the fractional abundances do not change very much.  We also calculated the column densities of CH$_3$OH with this method, and confirmed that these column densities are consistent with those determined by the multiple line analysis within a factor of 2.5.  In the following, we use the column densities of CH$_3$OH, HNCO and CS estimated by LTE approximation with $T_{\rm ex}$ of 5~K in order to discuss the abundances of various molecules on an equal footing.  

The H$^{13}$CN, H$^{13}$CO$^{+}$ and C$^{34}$S lines are detected in P1.  The isotope ratios are evaluated as: [HCN]/[H$^{13}$CN] = $25 \pm 18$, [HCO$^+$]/[H$^{13}$CO$^+$] = $30 \pm 28$, and [CS]/[C$^{34}$S] = $11 \pm 6$.  These values are lower than normal isotope ratios in the solar neighborhood: [$^{12}$C]/[$^{13}$C] = 59 and [$^{32}$S]/[$^{34}$S] = 22.  Hence, these normal spectral lines may be optically thick, if the solar neighborhood values of the isotope ratios could be applied to M51.  The [$^{18}$O]/[$^{17}$O] ratio is derived to be $13 \pm 6$ from the C$^{18}$O and C$^{17}$O data.  Since the lines of these two molecules are usually optically thin, the derived [$^{18}$O]/[$^{17}$O] ratio would be reliable.  The measured ratio is similar to or slightly higher than those observed in nuclear regions of external galaxies (6--10) \citep{sage1991, curran2001,wang2004,vila2008}, while it is higher than the Galactic value of 3--5.

HC$_3$N has been detected in many external galaxies \citep[\textit{e.g.}][]{henkel1988,aalto2002,aladro2011a,Lindberg2011}.  The fractional abundances of HC$_3$N averaged over the 1-kpc scale region in M51 are evaluated to be less than $(0.5-1.4)\times 10^{-9}$ in P1 and P2.  They are lower by a factor of two than those in IC~342 \citep[$X_{\rm HC_3N} \sim 1-2 \times 10^{-9}$:][]{meier2011}.  However, the IC~342 result is based on GMC-scale resolution (30~pc).  \citet{Lindberg2011} reported that the intensity ratios of HC$_3$N(10--9)/HCN(1--0) are 0.05--0.78 in the nuclear regions of external galaxies (resolution of 0.2 -- 24~kpc).  The ratio observed in M51~P1 is much lower ($<$ 0.02--0.04) than their values.  This result indicates that the abundance of HC$_3$N is lower in the spiral arm of M51 in the 1-kpc scale than in the nuclear regions.  The excitation conditions are likely not the reason for the lower intensity ratio, because the HNCO($6_{0\,6}$--$5_{0\,5}$) line which has a similar upper-state energy to that of the HC$_{3}$N(10--9) line, is observed as mentioned before.  If HC$_3$N is confined in a small dense region, the difference of the spatial resolution would result in different detectability of this molecule. 

The upper limits to deuterium fractionation ratios are found to be less than 0.8\% and 1.2~\% for DCO$^+$/HCO$^+$ and DCN/HCN, respectively.  Here, the column densities of HCN and HCO$^+$ are estimated from the normal species lines, because of the large uncertainty of the line intensities of the $^{13}$C isotopologues.  Therefore, the deuterium fractionation ratios would be smaller than these values, if the normal species lines are optically thick.  The upper limit to the DCN/HCN ratio is comparable to those reported for galactic star forming regions (Orion: 0.1-2~\%) \citep{Mangum1991,Schilke1992} and also cold dark clouds (1~\%) \citep{Turner2001}, while it is lower than those found in low-mass protostars (2--6~\%) \citep{Roberts2002}.  It is known that the deuterium fractionation ratios of molecules are as high as 10~\% for evolved starless cores without star formation \citep{Caselli2002}.  This is because they are enhanced in the later stage of prestellar cores due to a high degree of CO depletion onto dust grains.  The lower deuterium fractionation ratios in M51~P1 and P2 seem reasonable, because our single-dish observation samples less dense molecular clouds without heavy CO depletion occurring. 

\subsection{Line of Sight Velocities and Line Widths}
\begin{figure}
\epsscale{1.00}
\plotone{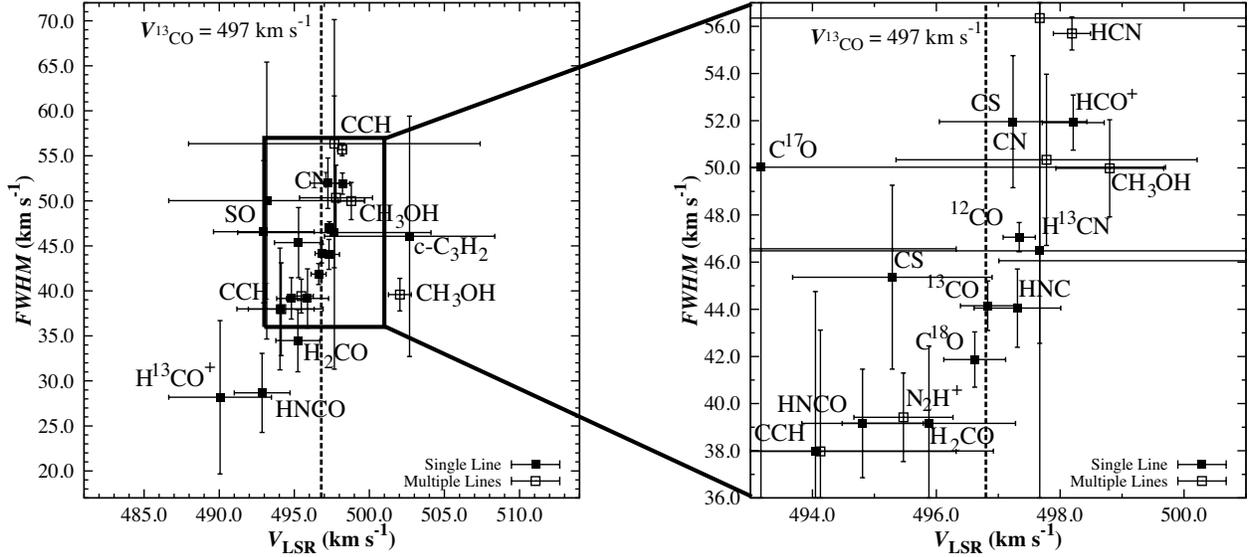}
\caption{Comparison of $V_{\rm LSR}$ and FWHM in M51~P1.  The right panel shows the expansion of the box in the left panel.  Filled squares indicate fitting results for a single line, whereas open squares indicate fitting results for lines with unresolved hyperfine or internal-rotation structure.  The error bars represents the fitting errors.  The vertical dashed-line indicates the line of sight velocity of $^{13}$CO. }
\label{fig4}
\end{figure}

Emitting regions could be different from molecule to molecule within the telescope beam.  In order to examine this, we prepared a correlation plot between the line of sight velocities and the line widths (FWHM) determined by Gaussian fitting to the spectra in P1 (Figure \ref{fig4}).  The lines of CO, HCO$^{+}$, and CS tend to have relatively broader line widths in comparison with the other molecular lines, probably because these molecules are widely distributed in the spiral arm.  The HCN, CCH, CN, and CH$_3$OH lines are broadened by blending of unresolved hyperfine components or internal rotation components.  The optical depth effect would also broaden the line width for CO, HCO$^{+}$, and CS.  In spite of these broadenings, molecular lines with relatively narrow line width tend to be blue-shifted relative to the velocity of the $^{12}$CO line, as shown in Figure \ref{fig4}.  

With the $\sim 0.''7$ resolution observation with CARMA, \citet{egusa11} found that massive GMA (giant molecular association) cores (GMC complexes) have blue-shifted velocities relative to the average velocity of the spiral arm.  These massive GMA cores are spatially located downstream of the spiral arm.  Therefore, molecules showing blue-shifted emissions could preferentially reside in the massive GMA cores identified by \citet{egusa11}.  If this is the case, the slight velocity differences may reflect different spatial distribution of molecular line emission within the beam.  This has to be confirmed by high resolution observations with interferometers.

\section{Discussion}
With this unbiased spectral line survey, we have revealed the chemical compositions averaged over the 1-kpc region in the spiral arm of M51.  This is the first systematic frequency scan toward a spiral arm in a nearby disk galaxy.

\subsection{Comparison between P1 and P2}
\begin{figure}
\epsscale{0.50}
\plotone{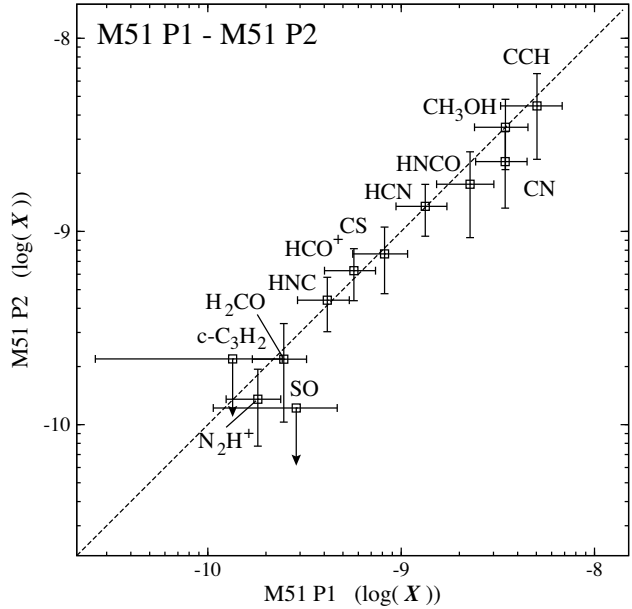}
\caption{Plots of the fractional abundances relative to H$_2$ between M51~P1 and P2. Arrows indicate upper limits.  The dashed line represents the equal abundances between P1 and P2.  A source size of 10$^{''}$ is assumed. }
\label{fig6}
\end{figure}

Figure \ref{fig6} shows a correlation plot of fractional abundances between the P1 and P2 positions, where a source size of $10^{''}$ is assumed.  The correlation plot does not change significantly with the assumed source size.  A good correlation of the abundances can be seen between the two positions. Although the abundance of CN is about 1.5 times higher in P1, most of the fractional abundance ratios between P1 and P2 are in a narrow range from 0.8 to 1.3.  When the uncertainty in each abundance is taken into account, the chemical composition can be regarded to be similar between P1 and P2.  On the other hand, star formation activities are higher in P1 than in P2, because the P1 position is bright in the H$\alpha$ and the 24~$\mu$m (Figure \ref{fig0}).  

In fact, the star formation rate (SFR) in P1 is estimated to be twice as high as in P2 on the basis of the 24~$\mu$m and H$\alpha$ data obtained by \citet{kennicutt2003}, where the SFRs are derived to be $0.055\pm 0.008$~M$_{\odot}$~yr$^{-1}$ and $0.022 \pm 0.004$~M$_{\odot}$~yr$^{-1}$ for P1 and P2, respectively, using the method by \citet{calzetti2007} (Appendix B).  The star formation efficiency, which is estimated from the SFR divided by the molecular gas mass, is 1.5 times higher in P1 ($(5.9 \pm 1.5) \times 10^{-10}$~yr$^{-1}$) than in P2 ($(3.9 \pm 1.0) \times 10^{-10}$~yr$^{-1}$), where the molecular gas mass is calculated to be $(9.4 \pm 2.0) \times 10^7$~M$_{\odot}$ and $(5.7 \pm 1.2) \times 10^7$~M$_{\odot}$ for P1 and P2, respectively, from the C$^{18}$O data obtained in this study.  

Hence, the star formation activity does not directly affect the chemical composition averaged over the 1-kpc scale region.  Although the chemical composition of star forming regions is known to be different from that in quiescent molecular clouds as observed in the Galactic sources, such as effect of star formation is diluted by the large observation beam.  Thus, the observed chemical composition primarily reflects that of the whole molecular clouds.  \citet{schinnere10} also discussed that star formation activities have no strong impact on the physical condition of molecular gas in the spiral arms at a linear resolution of 180~pc.  It is known that CCH and CN are deficient in high-mass star forming regions and are rather abundant in cloud envelopes \citep{Blake1987,Beuther2008}.  High abundances of these molecules are consistent with the picture that the observed chemical composition represents the GMC scale one.  The relatively high abundance of CH$_3$OH is, on the other hand, interesting, which is discussed in Section 4.3.

According to observational studies of nearby molecular cloud cores in the Galaxy, chemical compositions gradually changes along cloud evolution with a time scale of 10$^6$~yr.  If the dynamical time scale of clouds is less than this, a chemical evolution effect can be observed \citep[\textit{e.g.}][]{suzuki1992,Hirota2009}.  However, the dynamical time scale of 10~pc cloud, which is typical size of GMC, is roughly estimated to be a few $10^7$~yr, which corresponds to transit time of the cloud with the sound speed of the 0.3~km~s$^{-1}$, at the kinetic temperature of 15~K is \citep{Spitzer1978}.  Hence, such an chemical evolution effect will not be important for the GMC-scale chemical composition. 

\subsection{Comparison with Other Sources}
Chemical compositions reflect complex physical structures and conditions within the telescope beam.  Hence, it is generally difficult to discuss pure chemistry based on well-defined conditions, particularly for external galaxies.  Nevertheless, we can learn about the physical structure and conditions of the molecular gas from its chemical composition to some extent without resolving the source, because the different chemical compositions imply the different physical situations.  Hence, it is worth highlighting the chemical signatures specific to AGNs and starbursts by comparing their chemical compositions with those without such extreme activities (\textit{e.g.} those of M51~P1 and 2).  This is an important step for chemical diagnostics of external galaxies, and our spectral line survey toward M51~P1 and P2 is of fundamental importance in giving a reference template of molecular abundances in a spiral arm of a disk galaxy.  

\begin{figure}
\epsscale{1.00}
\plotone{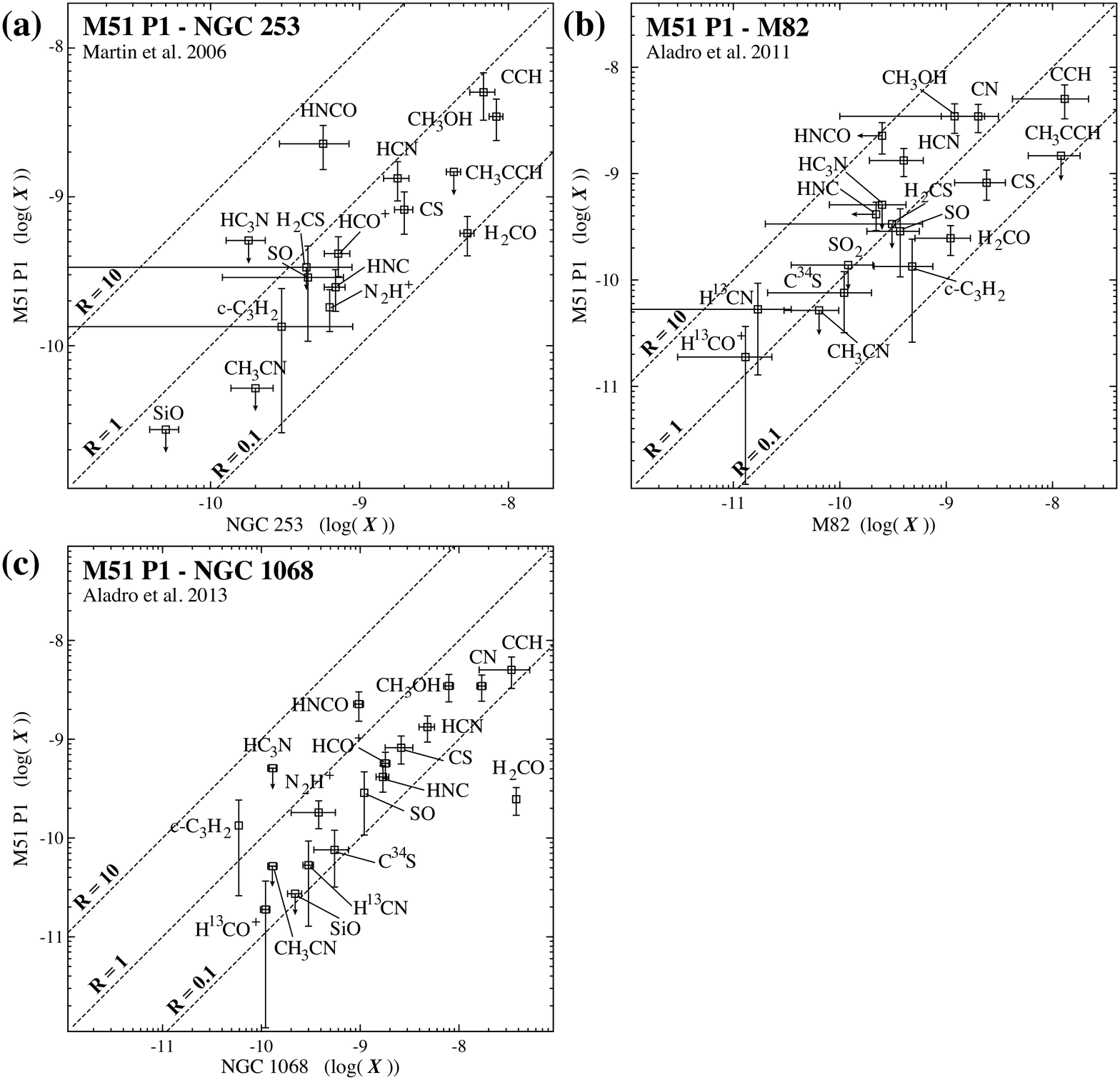}
\caption{Plots of the fractional abundances of molecules relative to H$_2$ between M51~P1 and other galaxies.  Arrows indicate upper limits.  R indicates the fractional abundance ratios of 10, 1, and 0.1 between M51~P1 and the other galaxies.  A source size of 10$^{''}$ is assumed for M51~P1.  The data for NGC~253, M82, and NGC~1068 are taken from \citet{martin2006}, \citet{aladro2011b}, and \citet{aladro2013}, respectively. } 
\label{fig7}
\end{figure}

For demonstrating how surrounding environments affect chemical compositions of GMCs, we compare the chemical composition of M51 P1 with those reported previously for the nuclear regions of other external galaxies with starbursts and AGNs.  Although these chemical compositions sample different size scales (0.3--1.5~kpc), we should be able to extract relations between such macroscopic chemical compositions and physical processes occurring therein.  Here, only M51~P1 is compared with other sources, because M51~P2 has a similar chemical composition to that of P1.  Figure \ref{fig7} shows correlation plots of the fractional abundances of M51~P1 against the starburst regions in NGC~253 and M82, as well as the AGN in NGC~1068.  For these sources, the fractional abundances are calculated by using the H$_2$ column density estimated from the C$^{18}$O data, assuming that [H$_2$]/[C$^{18}$O] = $2.9 \times 10^{6}$ \citep{meier05}.

Figures \ref{fig7} (a), (b), and (c) represent the results for NGC~253, M82, and NGC~1068, respectively.  Overall, molecular abundances of these sources correlate with that of M51 P1.  However, the correlations are loose in comparison with that between M51 P1 and P2.  This means that chemical composition of each source is different from that of M51 P1 because of an environmental effect.  Here, we highlight some characteristic features of each source below.

NGC~253 is a nearby \citep[3.4~Mpc:][]{Dalcanton2009} starburst galaxy and classified as SAB(s)c.  The SFR is estimated to be extremely high \citep[3.6~M$_{\odot}$~yr$^{-1}$:][]{Strickland2004} within a few hundred pc of the nuclear region.  \citet{martin2006} observed the central region of NGC~253 (($\alpha_{\rm J2000}$, $\delta_{\rm J2000}$) = (0$^{\rm h}$47$^{\rm m}$33$^{\rm s}$.3, -25$^{\circ}$17$^{'}$23$^{''}$)) with a resolution of 14--19$^{''}$ ($\sim$270~pc) using with the IRAM~30~m telescope in the 2~mm band.  They estimated column densities of detected molecules by using the rotation diagram method with a correction for beam dilution.  For molecules detected in a single transition, LTE with an excitation temperature of $T_{\rm ex} = 12$~K was assumed to estimate column densities.  In order to estimate the H$_2$ column density and the fractional abundances, we adapt the C$^{18}$O column density of $3.5 \times 10^{16}$~cm$^{-2}$ reported by \citet{Martin2010} on the basis of the observation of the C$^{18}$O($J=1-0$) line with the IRAM~30~m telescope which is derived under the LTE assumption with an excitation temperature of 10~K, without source size correction.  Although the C$^{18}$O column density is not corrected for the source size, the resultant error would be within a factor of 2.  By using the estimated fractional abundances of NGC~253, we compare the molecular abundances between M51~P1 and the starburst region in NGC~253 in the figure \ref{fig7} (a).  The abundance of HNCO tends to be lower in the starburst region of NGC~253 than in M51 P1, while the abundances of CS and H$_2$CO tend to be higher in NGC~253 than in M51 P1.  

M82 is also a nearby \citep[3.5~Mpc:][]{Karachentsev2004} edge-on galaxy with a starburst nucleus \citep[SFR $\sim9$~M$_{\odot}$~yr$^{-1}$:][]{Strickland2004}.  \citet{Fuente2008} reported that the chemical composition of the starburst region in M~82 is affected by photodissociation effects.  \citet{aladro2011b} observed the northeastern (NE) molecular lobe located at an offset position of ($\Delta\alpha$, $\Delta\delta$) = (+13.0$^{''}$, +7.5$^{''}$) from the dynamical center of ($\alpha_{\rm J2000}$, $\delta_{\rm J2000}$) = (9$^{\rm h}$55$^{\rm m}$51$^{\rm s}$.9, +69$^{\circ}$40$^{'}$47$^{''}$.10) with the IRAM~30~m telescope in the 2~mm and 1.3~mm bands.  The resolution of 19--14$^{''}$ in the 2~mm band and that of 10--9$^{''}$ in the 1.3~mm band correspond to $\sim$280~pc and $\sim$160~pc, respectively.  In addition to their observational data, they complied observational data of C$^{18}$O \citep{martin2009}, CN and HCN \citep{Fuente2005}, HNC \citep{huettemeister1995}.  They estimated the column densities of detected molecules by using the rotation diagram method with a correction for beam dilution.  For molecules detected in a single transition, they assumed the LTE with $T_{\rm ex} = 20 \pm 10$~K to estimate the column densities.  By using the abundances reported by \citet{aladro2011b}, we compare the fractional abundances between M51~P1 and M~82 in figure \ref{fig7} (b).  M~82 has a lower abundance of HNCO, as well as higher abundance of CS and H$_2$CO than those in M51~P1 similar to the trends found in NGC~253.  

NGC~1068 is a nearby \citep[14.4~Mpc;][]{Bland-Hawthorn1997} Seyfert~2 galaxy.  The chemical composition of the nuclear region suggests that a giant X-ray-dominated region is present in NGC~1068 \citep[\textit{e.g.}][]{Kohno2001,usero2004,Krips2008}.  \citet{aladro2013} observed the nuclear region of NGC~1068 with resolution of 21--29$^{''}$ (1.5--2~kpc) using the IRAM~30~m telescope in the 3~mm band.  They estimated the column densities of the detected molecules via the rotation diagram method in combination with other transition data taken from the other literatures \citep{usero2004, Krips2008, Bayet2009, Perez2009, nakajima2011}.  For molecules detected in a single transition, they calculated the column densities under the assumption of LTE with $T_{\rm ex}$ of $10 \pm 5$~K.  Figure \ref{fig7} (c) shows the comparison of fractional abundances between M51~P1 and the AGN region in NGC~1068 \citep{aladro2013}.  In addition to the similar chemical characteristics found in M82, NGC~1068 has higher abundances of SiO, HCN, CN, and CCH.  The enhancement of SiO and HCN is reported in the circumnuclear gas around the AGN where strong X-ray irradiation from the AGN is expected \citep{usero2004, garcia2010}.  

Deficiency of HNCO and enhancement of H$_2$CO, and CS are seen in the starburst regions and the AGN in comparison with M51 P1.  \citet{martin2009} reported that HNCO/CS abundance ratios vary by about two orders of magnitude among the starburst galaxies.  In photodissociation regions (PDR), CS is thought to be formed efficiently due to abundant S$^+$ which is ionized from S by UV photons \citep{Drdla1989}.  Therefore, the strong UV radiation field in the nuclear regions seems to be consistent with the observed trend.  Indeed, the UV radiation field from embedded young massive stars, is expected to be much higher in the starburst galaxies than M51~P1, because the SFRs are about two orders of magnitude higher in the starburst regions of these galaxies.  On the other hand, the lower HNCO abundance in the nuclear regions is puzzling.  Although the production mechanism of HNCO is still controversial, HNCO is often thought to be a shock tracer \citep[\textit{e.g.}][]{rodriguez2010}.  Strong shocks, which is caused, for instance, by accretion shock of inflowing gas to the nuclear region, star formation feedbacks, and jet from AGN, are expected in the nuclear regions.  Therefore, the lower abundance of HNCO in the nuclear regions means that strong UV or X-ray may rather affect its formation and destruction chemistry.  This is left for future studies.  

\subsection{Abundance of CH$_3$OH }
In this survey, CH$_3$OH is found to be relatively abundant toward both P1 and P2.  It is more abundant than HCN, HCO$^+$, and CS, which are widely observed in external galaxies.  Although the fractional abundances of HCN, HCO$^+$, and CS may be underestimated due to the optical depth effect of their emission lines, the underestimation is expected to be less than a factor of 3 based on the H$^{13}$CN, H$^{13}$CO$^{+}$, and C$^{34}$S intensities.  Thus, the abundance of CH$_3$OH is significantly higher than those of HCN, HCO$^+$, and CS.  Furthermore, the CH$_3$OH abundances in M51~P1 and P2 are comparable to those found in the nuclear regions as shown in Figure \ref{fig7}.  CH$_3$OH enhancement is often discussed in terms of shock excitation in the nuclear regions \citep[\textit{e.g.}][]{garcia2010} and the bar regions\citep{meier05,meier2012,usero2006}.  If shocks are principal cause of CH$_3$OH enhancement in the M51 case, spiral shocks \citep[\textit{e.g.}][]{Roberts1969} and cloud-cloud collisions are possible mechanism of shocks in the spiral arm.  

Alternatively, it is worth considering non-thermal desorption of CH$_3$OH as a possibility.
The main formation process of CH$_3$OH is considered to be hydrogenation of the CO molecule on dust grains under a low temperature condition \citep{tielens1997,watanabe2003}.  Hence, high temperature conditions ($T>20$~K) or energetic phenomena are required for liberation of CH$_3$OH into the gas phase.  For instance, CH$_3$OH has been detected in shocked regions caused by an impact of outflows from protostars \citep[\textit{e.g.}][]{bachiller1997,Sakai2012}, and it is often regarded as a shock tracer as mentioned above.  Nevertheless, CH$_3$OH exists in the cold widespread gas of M51~P1 and P2 observed with our telescope beam.  Hence, its origin is of great interest.  We examine the required liberation rate to account for the observed CH$_3$OH abundances in the gas phase.

For simplicity, we consider evaporation, depletion, and gas-phase destruction with ions.  The rate equation for CH$_3$OH is approximately written as:
\begin{equation}
\frac{d[{\rm CH_3OH}]}{dt} = n_{\rm g} k_{\rm ev} - k_{\rm d} [{\rm CH_3OH}] - k_{\rm r}[{\rm CH_3OH}][{\rm M^+}],
\label{eq1}
\end{equation}
where $k_{\rm ev}$ is the liberation rate coefficient of CH$_3$OH per dust grain, $n_{\rm g}$ is the number density of dust grains, $k_{\rm d}$ is the rate coefficient of depletion of CH$_3$OH onto dust grain, and $k_{\rm r}$ is the total rate coefficient for reactions of CH$_3$OH with various ions M$^+$ (M$^+$ = H$^+$, H$_3^+$, HCO$^+$, and He$^+$).  Assuming the steady-state condition, equation (\ref{eq1}) is modified to:
\begin{equation}
k_{\rm ev} = \frac{X_{\rm CH_3OH}(k_{\rm d} + k_{\rm r} [{\rm e}])}{X_{\rm g}}.
\end{equation}
Here, we approximately assume that $[{\rm M^+}] \sim [{\rm e}]$, where $[{\rm e}]$ is a number density of electrons in the gas phase.  Furthermore, $X_{\rm CH_3OH}$ and  $X_{\rm d}$ represent fractional abundances of CH$_3$OH and grains, respectively, relative to H$_2$.   
For the H$_2$ density of $10^3$~cm$^{-3}$, the following parameters are assumed: $k_{\rm d} \sim n({\rm H_2}) / 3 \times 10^{16}$~s$^{-1}$ $\sim 3 \times 10^{-14}$~s$^{-1}$ \citep{caselli1999}, $k_{\rm r} \sim 10^{-9}$~cm$^{3}$~s$^{-1}$ (a typical Langevin rate coefficient), $[{\rm e}] \sim 10^{-5} \times \sqrt{n({\rm H_2})}$~cm$^{-3}$ $\sim 3 \times 10^{-4}$~cm$^{-3}$, and $X_{\rm g} \sim 10^{-12}$ \citep{hasegawa1992}.  Then, $k_{\rm ev}$ is required to be $\sim 1 \times 10^{-9}$~s$^{-1}$ in order to reproduce the observed value of $X_{\rm CH_3OH}$ of $3.1 \times 10^{-9}$.  When the H$_2$ density of $10^4$~cm$^{-3}$ is assumed, $k_{\rm ev}$ has to be $\sim 4 \times 10^{-9}$~s$^{-1}$.


This liberation rate can be compared with that in the nearby quiescent starless core L134N.  In this source, the fractional abundance of CH$_3$OH is $5.07 \times 10^{-9}$ \citep{Dickens2000}.  Hence, the liberation rate is evaluated to be $5 \times 10^{-9}$~s$^{-1}$ for an H$_2$ density of $3 \times 10^{4}$~cm$^{-3}$ by using equation (4).  This value is comparable to that found in M51~P1, although the spatial scale is much different.  

The liberation mechanism of CH$_3$OH is unclear.  In order to thermally desorb the CH$_3$OH molecules from dust grains, the temperature has to be as high as 100~K.  In the spiral arm of M51, the star formation activity is not a potential mechanism, because the star formation feedback seems to be localized to small areas.  Indeed, our survey shows no enhancement of the CH$_3$OH fractional abundance in P1 which shows higher star formation activities than in P2.  A possible mechanism is a large-scale shock such as spiral shocks and cloud-cloud collisions.  Indeed, an enhancement of CH$_3$OH has been observed in the central spiral arm regions of IC~342 \citep{meier05,usero2006} and Maffei~2 \citep{meier2012}, which are confirmed as shocked regions by observations of the shock tracer SiO.  The abundances of CH$_3$OH are estimated to be $(3-8) \times 10^{-9}$ and $(3 - 20) \times 10^{-9}$ in IC~342 and Maffei~2 \citep{meier05,meier2012}, respectively.  These abundances are similar to or higher than those of M51, although the spatial resolution of the IC~342 and Maffei~2 observation is much higher (50 -- 120~pc) than that of our M51 observation.  Alternatively, non-thermal desorption process such as photoevaporation by cosmic-ray induced UV photons \citep[\textit{e.g.}][]{Prasad1983} could be responsible for evaporation of CH$_3$OH. Since L134N does not show any evidence of shocks, the origin of CH$_3$OH is likely the non- thermal desorption process.  If the cosmic ray flux in M51~P1 is comparable to that in L134N, this mechanism can explain the observed abundance of CH$_3$OH.  In any case, it seems important to explore the distribution of CH$_3$OH in M51~P1 with higher angular resolution in order to assess the evaporation mechanism.

\subsection{Summary}
We carried out a spectral line survey toward two positions (P1 and P2) in the spiral arm of the nearby galaxy M51 in the 3~mm and 2~mm bands with the IRAM 30~m telescope.  This is the first systematic frequency-scan toward a spiral arm of a nearby galaxy, and the result can be used as a template for the 1-kpc scale chemical composition of a spiral arm.  Main results are summarized as follows:

\begin{enumerate}
\item Thirteen molecular species, CCH, CN, HCN, HNC, CO, HCO$^+$, H$_2$CO, CH$_3$OH, N$_2$H$^+$, c-C$_3$H$_2$, HNCO, CS, and SO were identified in P1.  In addition, six isotopologues, $^{13}$CO, C$^{17}$O , C$^{18}$O, H$^{13}$CO$^+$, H$^{13}$CN, and C$^{34}$S were identified.  Eleven molecular species and three CO isotopologues were found in P2.
\item HC$_3$N, CH$_3$CN, CH$_3$CCH, and SiO, which are often detected in the nuclear regions of nearby galaxies, were not detected in the both observed positions.  The upper limit to the column density of HC$_3$N is much lower than those found in nuclear regions such as M82's one.
\item Deuterated species such as DCN and DCO$^+$ were not detected in this survey, either.  The upper limits to the deuterium fractionation ratios are comparable to those in Galactic star-forming cores and young dark clouds (TMC-1), whereas they are lower than those in low-mass star forming regions. 
\item P1 and P2 are found to have almost identical chemical composition, although the star formation efficiency in P1 is 1.5 times higher than that in P2.  Therefore, a small difference of star formation activities has little influence on the chemical composition averaged over the 1-kpc scale region in the spiral arm.
\item The chemical compositions of M51~P1 is found to be loosely correlated with those of starbursts and AGNs.  The difference of chemical compositions between M51 P1 and the nuclear regions, such as enhancement of CS in the nuclear regions may be attributed to the extreme environments such as high UV or X-ray radiation fields in the nuclear regions.
\item CH$_3$OH is found to be abundant in the molecular gas in the spiral arm.  Because CH$_3$OH is thought to be formed on the dust mantle, some liberation mechanisms are necessary.  A shock induced by kpc-scale dynamics such as cloud-cloud collisions, as well as non-thermal desorption processes, could be potential mechanisms. 
\end{enumerate}

\acknowledgments

The authors are grateful to the IRAM staff for excellent support.  This study is supported by a Grant-in-Aid from the Ministry of Education, Culture, Sports, Science, and Technology of Japan (No. 21224002, 21740132, and 25108005).  




\clearpage


\clearpage


\clearpage
\begin{deluxetable}{llllll}
\tablecolumns{6}
\tablewidth{0pt}
\tablecaption{Frequency Settings }
\tablehead{
\multicolumn{6}{c}{M51~P1}\\
\colhead{ } & \colhead{Integ. $^{\rm a}$ (min)} & \colhead{Reciever $^{\rm b}$ } & \colhead{LSB (GHz)} & \colhead{USB (GHz)} 
& \colhead{$T_{\rm sys}$ (K)} }
\startdata
Set 1 & 219 & E090 & ~82.000 - ~90.480 & --                & ~75 - 120 \\
      &     & E150 & 144.250 - 148.300 & --                & ~80 - 110 \\
Set 2 & 173 & E090 & ~90.200 - ~97.980 & --                & ~90 - 130 \\
      &     & E150 & 138.000 - 142.050 & --                & ~90 - 100 \\
Set 3 & 260 & E090 & ~90.200 - ~97.980 & --                & ~80 - 120 \\
      &     & E150 & 142.000 - 146.050 & --                & ~85 - 130 \\
Set 4 & 159 & E090 & ~97.700 - 105.580 & --                & ~80 - 110 \\
      &     & E150 & 130.000 - 134.050 & --                & 100 - 120 \\
Set 5 & 188 & E090 & ~97.700 - 105.580 & --                & 100 - 130 \\
      &     & E150 & 134.000 - 138.050 & --                & ~80 - 110 \\
Set 6 & 173 & E090 & ~92.720 - 100.500 & 108.420 - 116.200 & 140 - 300 \\
      &     & E150 & --                & --                & --  \\
\cutinhead{M51~P2}
Set 1 & 143 & E090 & ~85.020 - ~92.800 & 100.720 - 108.500 & 110 - 290 \\
      &     & E150 & --                & --                & --  \\
Set 2 & 360 & E090 & ~92.720 - 100.500 & 108.420 - 116.200 & 110 - 310 \\
      &     & E150 & --                & --                & --  \\
Set 3 & 298 & E090 & ~85.020 - ~92.800 & --                & 110 - 170  \\
      &     & E150 & 139.250 - 143.300 & --                & 150 - 210 \\
\enddata
\tablenotetext{a}{Total on-source integration time.}
\tablenotetext{b}{EMIR receiver names.}
\tablenotetext{c}{Typical range of the system noise temperature during the observation.}
\label{tab7}
\end{deluxetable}

\clearpage
\begin{deluxetable}{lrlrrrrrr}
\rotate
\tabletypesize{\footnotesize}
\tablecolumns{9}
\tablewidth{620pt}
\tablecaption{Line parameters observed in M51~P1}
\tablehead{
\colhead{Name} & \colhead{Frequency} & \colhead{Transition} & \colhead{$E_{\rm u}$} & \colhead{S$\mu^2$} & \colhead{$T_{\rm mb}$ Peak \tablenotemark{a}} & \colhead{$\int T_{\rm mb} dv$ \tablenotemark{a}} & \colhead{$V_{\rm LSR}$} & \colhead{FWHM} \\
\colhead{} & \colhead{(GHz)} & \colhead{} & \colhead{(K)} & \colhead{} & \colhead{(mK)} & \colhead{(K~km~s$^{-1}$)} & \colhead{(km~s$^{-1}$)} &\colhead{(km~s$^{-1}$)} 
}
\startdata
OCS            & 85.139103 &$7-6$                       & 16.3 &  3.58 & $< 2$ & $< 0.05$        &                 &               \\
c-C$_3$H$_2$   & 85.338894 &$2_{1\,2}-1_{0\,1}$         &  6.4 &  16.1 & $2 \pm 2$ & $0.08 \pm 0.06$ & $503 \pm 6$ & $46 \pm 13$  \\
CH$_3$CCH      & 85.457300 &$5_0-4_0$                   & 12.3 &  6.15 & $< 2$ & $< 0.04$        &                 &  \\
H$^{13}$CN     & 86.339922 &$1-0$                       &  4.1 &  8.91 & $3 \pm 2$ & $0.10 \pm 0.07$ & $498 \pm 6$ & $46 \pm 15$  \\
H$^{13}$CO$^+$ & 86.754288 &$1-0$                       &  4.2 &  15.2 & $2 \pm 2$ & $0.06 \pm 0.06$ & $490 \pm 3$ & $28 \pm 9$  \\
SiO            & 86.846960 &$2-1$                       &  6.3 &  19.2 & $< 2$ & $< 0.04$        &                 &  \\
HN$^{13}$C     & 87.090825 &$1-0$                       &  4.2 &  9.30 & $< 2$ & $< 0.05$        &                 &  \\
CCH            & 87.316898 &$N=1-0,J=5/2-3/2,F=2-1$     &  4.2 &  0.99 & $5 \pm 2$ & $0.41 \pm 0.08$ & $498 \pm 10$ & $56 \pm 14$ \\
CCH            & 87.328585 &$N=1-0,J=5/2-3/2,F=1-0$     &  4.2 &  0.49 & --    &              -- &              -- & -- \\ 
CCH            & 87.401989 &$N=1-0,J=3/2-3/2,F=1-1$     &  4.2 &  0.49 & $4 \pm 2$ & $0.16 \pm 0.06$ & $494 \pm 2$ & $38 \pm 5$ \\
CCH            & 87.407165 &$N=1-0,J=3/2-3/2,F=0-1$     &  4.2 &  0.20 & --    &              -- &              -- & --  \\
HNCO           & 87.925237 &$4_{0\,4}-3_{0\,3}$         & 10.6 &  10.0 & $9 \pm 2$ & $0.40 \pm 0.06$ & $495 \pm 1$ & $39 \pm 2$ \\
HCN            & 88.631602 &$1-0$                       &  4.3 &  8.91 & $45 \pm 2$ & $2.74 \pm 0.09$ & $498.2 \pm 0.3$ & $55.7 \pm 0.7$ \\
HCO$^+$        & 89.188525 &$1-0$                       &  4.3 &  15.2 & $37 \pm 2$ & $2.04 \pm 0.08$ & $498.2 \pm 0.5$ & $52 \pm 1$ \\
HNC            & 90.663568 &$1-0$                       &  4.4 &  9.30 & $19 \pm 1$ & $0.96 \pm 0.06$ & $497.3 \pm 0.7$ & $44 \pm 2$ \\
HC$_3$N        & 90.979023 &$10-9$                      & 24.0 & 139.3 & $< 2$ & $<  0.04$       &                 &  \\
CH$_3$CN       & 91.987088 &$5_0-4_0$                   & 13.2 &  76.9 & $< 2$ & $<  0.04$       &                 &  \\
N$_2$H$^+$     & 93.173392 &$1-0$                       &  4.5 & 111.8 & $12 \pm 2$ & $0.53 \pm 0.06$ & $495.5 \pm 0.8$ & $39 \pm 2$ \\
C$^{34}$S      & 96.412950 &$2-1$                       &  6.9 &  7.67 & $ 2 \pm 1$ & $0.06 \pm 0.03$ & $500 \pm 3$ & $30 \pm 6$  \\
CH$_3$OH       & 96.741375 &$2_0-1_0,{\rm A}^+$         &  7.0 &  1.62 & $11 \pm 2$ & $0.52 \pm 0.05$ & $498.8 \pm 0.9$ & $50 \pm 2$ \\
CH$_3$OH       & 96.739362 &$2_{-1}-1_{-1},{E}$         &  4.6 &  1.21 & --   &              -- &              -- & -- \\
CH$_3$OH       & 96.744550 &$2_{0}-1_{0},{E}$           & 13.6 &  1.62 & --   &              -- &              -- & -- \\
CS             & 97.980953 &$2-1$                       &  7.1 &  7.67 & $13 \pm 2$ & $0.68 \pm 0.08$ & $497 \pm 1$ & $52 \pm 3$ \\
SO             & 99.299870 &$N_J=2_3-1_2$               &  9.2 &  6.91 & $2 \pm 2$ & $0.09 \pm 0.05$ & $493 \pm 3$ & $47 \pm 8$ \\
HC$_3$N        &100.076392 &$11-10$                     & 28.8 & 153.2 & $< 2$ & $< 0.04 $       &                 &  \\
H$_2$CS        &103.040452 &$3_{0\,3} - 2_{0\,2}$       &  9.9 &  8.16 & $< 2$ & $< 0.03 $       &                 &  \\
SO$_2$         &104.029418 &$3_{1\,3} - 2_{0\,2}$       &  7.7 &  5.37 & $< 2$ & $< 0.03 $       &                 &  \\
C$^{18}$O      &109.782173 &$1-0$                       &  5.3 & 0.012 & $42 \pm 4$ & $1.9 \pm 0.1$ & $496.6 \pm 0.5$ & $42 \pm 2$ \\
HNCO           &109.905749 &$5_{0\,5}-4_{0\,4}$         & 15.8 &  12.5 & $10 \pm 3$ & $0.3 \pm 0.1$ & $494 \pm 3$ & $38 \pm 7$ \\  
$^{13}$CO      &110.201354 &$1-0$                       &  5.3 & 0.012 & $158 \pm 4$ & $7.5 \pm 0.2$ & $496.8 \pm 0.4$ & $44 \pm 1$ \\
C$^{17}$O      &112.359284 &$1-0$                       &  5.4 & 0.012 & $6 \pm 4$ & $0.16 \pm 0.05$ & $493 \pm 7$ & $50 \pm 15$ \\
CN             &113.144157 &$N=1-0,J=1/2-1/2,F=1/2-3/2$ &  5.4 &  1.25 & $9 \pm 6$ & $1.21 \pm 0.06$ & &  \\
CN             &113.170492 &$N=1-0,J=1/2-1/2,F=3/2-1/2$ &  5.4 &  1.22 & --    &              -- &              -- & -- \\
CN             &113.191279 &$N=1-0,J=1/2-1/2,F=3/2-3/2$ &  5.4 &  1.58 & --    &              -- &              -- & -- \\
CN             &113.490970 &$N=1-0,J=3/2-1/2,F=5/2-3/2$ &  5.4 &  4.21 & $28 \pm 5$ & $2.08 \pm 0.05$ & $498 \pm 2$ & $50 \pm 4$ \\
CN             &113.488120 &$N=1-0,J=3/2-1/2,F=3/2-1/2$ &  5.4 &  1.58 & --    &              -- &              -- & -- \\
CN             &113.499644 &$N=1-0,J=3/2-1/2,F=1/2-1/2$ &  5.4 &  1.25 & --    &              -- &              -- & -- \\
$^{12}$CO      &115.271202 &$1-0$                       &  5.5 & 0.012 & $1378 \pm 7$ & $68.6 \pm 0.3$  & $497.3 \pm 0.3$ & $47.1 \pm 0.6$ \\
SiO            &130.268610 &$3-2$                       & 12.5 &  28.8 & $< 3$ & $< 0.07 $       &                 &  \\
HNCO           &131.885734 &$6_{0\,6}-5_{0\,5}$         & 22.2 &  15.0 & $6 \pm 2$ & $0.22 \pm 0.08$ & $492 \pm 2$ & $29 \pm 4$ \\
OCS            &133.785900 &$11-10$                     & 38.5 &  5.63 & $< 3$ & $< 0.06$        &                 &               \\
HDCO           &134.284830 &$2_{1\,1}-1_{1\,0}$         & 17.6 &  8.15 & $< 2$ & $< 0.04$        &                 &  \\
SO$_2$         &135.696020 &$5_{1\,5} - 4_{0\,4}$       & 15.7 &  8.35 & $< 2$ & $< 0.04$        &                 &  \\
HC$_3$N        &136.464411 &$15-14$                     & 52.4 & 208.9 & $< 2$ & $< 0.04$        &                 &  \\
CH$_3$CCH      &136.728012 &$8_0-7_0$                   & 29.5 &  9.83 & $< 2$ & $< 0.04$        &                 &  \\
H$_2$CS        &137.371210 &$4_{0\,4} - 3_{0\,3}$       & 16.5 &  10.9 & $< 2$ & $< 0.04$        &                 &  \\
SO             &138.178600 &$N_J=3_4-2_3$               & 15.9 &  9.28 & $< 3$ & $< 0.05$        &                 &  \\
H$_2$CO        &140.839502 &$2_{1\,2}-1_{1\,1}$         & 21.9 &  8.16 & $8 \pm 2$ & $0.33 \pm 0.06$ & $496 \pm 1$ & $39 \pm 3$ \\
H$_2^{13}$CO   &141.983740 &$2_{0\,2}-1_{0\,1}$         & 10.2 &  10.9 & $< 2$ & $< 0.04$        &                 &  \\
DCO$^+$        &144.077289 &$2-1$                       & 10.4 &  30.4 & $< 2$ & $< 0.03$        &                 &  \\
DCN            &144.828002 &$2-1$                       & 10.4 &  17.9 & $< 2$ & $< 0.03$        &                 &  \\
CH$_3$OH       &145.093707 &$3_0-2_0,{\rm E}$           & 19.2 &  2.42 & $14 \pm 2$ & $0.60 \pm 0.06$ & $502.0 \pm 0.7$ & $40 \pm 2$ \\
CH$_3$OH       &145.097370 &$3_{-1}-2_{-1},{\rm E}$     & 11.6 &  2.16 & --    &              -- &              -- & -- \\
CH$_3$OH       &145.103152 &$3_0-2_0,{\rm A}^+$         & 13.9 &  2.43 & --    &              -- &              -- & -- \\
H$_2$CO        &145.602949 &$2_{0\,2}-1_{0\,1}$         & 10.5 &  10.9 & $4 \pm 1$ & $0.14 \pm 0.03$ & $495 \pm 1$ & $34 \pm 3$ \\
CS             &146.969029 &$3-2$                       & 14.1 &  11.5 & $9 \pm 3$ & $0.4 \pm 0.1$ & $495 \pm 2$ & $45 \pm 4$ \\
CH$_3$CN       &147.174588 &$8_0-7_0$                   & 31.8 & 246.1 & $< 3$ & $<  0.05$       &                 &  \\
\enddata
\tablenotetext{a}{The errors are 3$\sigma$.  The upper limit to the integrated intensity is calculated as: $\int T_{\rm mb} dv < 3 \sigma \times \sqrt{\Delta V / \Delta v_{\rm res}} \Delta v_{\rm res}$, where $\Delta V$ is the assumed line width (40~km~s$^{-1}$) and $\Delta v_{\rm res}$ is the velocity resolution per channel.}
\label{tab1}
\end{deluxetable}

\clearpage

\begin{deluxetable}{lrlrrrrrr}
\rotate
\tabletypesize{\footnotesize}
\tablecolumns{9}
\tablewidth{620pt}
\tablecaption{Line parameters observed in M51~P2}
\tablehead{
\colhead{Name} & \colhead{Frequency} & \colhead{Transition} & \colhead{$E_{\rm u}$} & \colhead{S$\mu^2$} & \colhead{$T_{\rm mb}$ Peak \tablenotemark{a}} & \colhead{$\int T_{\rm mb} dv$ \tablenotemark{a}} & \colhead{$V_{\rm LSR}$} & \colhead{FWHM} \\
\colhead{} & \colhead{(GHz)} & \colhead{} & \colhead{(K)} & \colhead{} & \colhead{(mK)} & \colhead{(K~km~s$^{-1}$)} & \colhead{(km~s$^{-1}$)} &\colhead{(km~s$^{-1}$)} 
}
\startdata
OCS            & 85.139103 &$7-6$                       & 16.3 &  3.58 & $< 2$ & $< 0.05$ &  & \\
c-C$_3$H$_2$   & 85.338894 &$2_{1\,2}-1_{0\,1}$         &  6.4 &  16.1 & $< 2$ & $< 0.06$ &  & \\
CH$_3$CCH      & 85.457300 &$5_0-4_0$                   & 12.3 &  6.15 & $< 2$ & $< 0.05$ &  & \\
H$^{13}$CN     & 86.339922 &$1-0$                       &  4.1 &  8.91 & $< 2$ & $< 0.05$ &  & \\
H$^{13}$CO$^+$ & 86.754288 &$1-0$                       &  4.2 &  15.2 & $< 2$ & $< 0.06$ &  & \\
SiO            & 86.846960 &$2-1$                       &  6.3 &  19.2 & $< 2$ & $< 0.04$ &  & \\
HN$^{13}$C     & 87.090825 &$1-0$                       &  4.2 &  9.30 & $< 2$ & $< 0.05$ &  & \\
CCH            & 87.316898 &$N=1-0,J=5/2-3/2,F=2-1$     &  4.2 &  0.99 & $4 \pm 2$ & $0.22 \pm 0.08$ & $503 \pm 7$ & $77 \pm 16$ \\
CCH            & 87.328585 &$N=1-0,J=5/2-3/2,F=1-0$     &  4.2 &  0.49 & --    &              -- &              -- & -- \\
CCH            & 87.401989 &$N=1-0,J=3/2-3/2,F=1-1$     &  4.2 &  0.49 & $2 \pm 2$ & $0.16 \pm 0.09$ & \\
CCH            & 87.407165 &$N=1-0,J=3/2-3/2,F=0-1$     &  4.2 &  0.20 & --    &              -- &              -- & --  \\
HNCO           & 87.925237 &$4_{0\,4}-3_{0\,3}$         & 10.6 &  10.0 & $4 \pm 2$ & $0.19 \pm 0.07$ & $507 \pm 4$ & $43 \pm 9$ \\
HCN            & 88.631602 &$1-0$                       &  4.3 &  8.91 & $29 \pm 2$ & $1.7 \pm 0.1$ & $516.0 \pm 0.9$ & $54 \pm 2$ \\
HCO$^+$        & 89.188525 &$1-0$                       &  4.3 &  15.2 & $24 \pm 2$ & $1.38 \pm 0.09$ & $514.8 \pm 0.9$ & $52 \pm 1$ \\
HNC            & 90.663568 &$1-0$                       &  4.4 &  9.30 & $11 \pm 2$ & $0.63 \pm 0.07$ & $514 \pm 1$ & $44 \pm 2$ \\
HC$_3$N        & 90.979023 &$10-9$                      & 24.0 & 139.3 & $< 2$ & $<  0.05$       &                 &  \\
CH$_3$CN       & 91.987088 &$5_0-4_0$                   & 13.2 &  76.9 & $< 2$ & $<  0.05$       &                 &  \\
N$_2$H$^+$     & 93.173392 &$1-0$                       &  4.5 & 111.8 & $6 \pm 3$ & $0.24 \pm 0.08$ & $511 \pm 3$ & $44 \pm 7$ \\
C$^{34}$S      & 96.412950 &$2-1$                       &  6.9 &  7.67 & $< 2$ & $< 0.06$ &  & \\
CH$_3$OH       & 96.741375 &$2_0-1_0,{\rm A}^+$         &  7.0 &  1.62 & $8 \pm 2$ & $0.32 \pm 0.09$ & $513 \pm 1$ & $38 \pm 2$ \\
CH$_3$OH       & 96.739362 &$2_{-1}-1_{-1},{E}$         &  4.6 &  1.21 & --   &              -- &              -- & -- \\
CH$_3$OH       & 96.744550 &$2_{0}-1_{0},{E}$           & 13.6 &  1.62 & --   &              -- &              -- & -- \\
CS             & 97.980953 &$2-1$                       &  7.1 &  7.67 & $7 \pm 2$ & $0.39 \pm 0.09$ & $518 \pm 1$ & $49 \pm 3$ \\
SO             & 99.299870 &$N_J=2_3-1_2$               &  9.2 &  6.91 & $< 2$ & $< 0.05$ &  & \\
HC$_3$N        &100.076392 &$11-10$                     & 28.8 & 153.2 & $< 2$ & $< 0.05 $       &                 &  \\
H$_2$CS        &103.040452 &$3_{0\,3} - 2_{0\,2}$       &  9.9 &  8.16 & $< 4$ & $< 0.08 $       &                 &  \\
SO$_2$         &104.029418 &$3_{1\,3} - 2_{0\,2}$       &  7.7 &  5.37 & $< 4$ & $< 0.09 $       &                 &  \\
C$^{18}$O      &109.782173 &$1-0$                       &  5.3 & 0.012 & $25 \pm 3$ & $1.16 \pm 0.09$ & $517.4 \pm 0.6$ & $43 \pm 1$ \\
HNCO           &109.905749 &$5_{0\,5}-4_{0\,4}$         & 15.8 &  12.5 & $4 \pm 3$ & $0.17 \pm 0.07$ & $499 \pm 2$ & $35 \pm 5$ \\  
$^{13}$CO      &110.201354 &$1-0$                       &  5.3 & 0.012 & $109 \pm 2$ & $5.2 \pm 0.1$ & $519.1 \pm 0.3$ & $44.2 \pm 0.7$ \\
C$^{17}$O      &112.359284 &$1-0$                       &  5.4 & 0.012 & $4 \pm 3$ & $0.13 \pm 0.08$ & $519 \pm 5$ & $45 \pm 12$ \\
CN             &113.144157 &$N=1-0,J=1/2-1/2,F=1/2-3/2$ &  5.4 &  1.25 & $5 \pm 3$ & $0.5 \pm 0.2$ &  & \\
CN             &113.170492 &$N=1-0,J=1/2-1/2,F=3/2-1/2$ &  5.4 &  1.22 & --    &              -- &              -- & -- \\
CN             &113.191279 &$N=1-0,J=1/2-1/2,F=3/2-3/2$ &  5.4 &  1.58 & --    &              -- &              -- & -- \\
CN             &113.490970 &$N=1-0,J=3/2-1/2,F=5/2-3/2$ &  5.4 &  4.21 & $10 \pm 4$ & $0.9 \pm 0.2$ & $509 \pm 4$ & $46 \pm 6$ \\
CN             &113.488120 &$N=1-0,J=3/2-1/2,F=3/2-1/2$ &  5.4 &  1.58 & --    &              -- &              -- & -- \\
CN             &113.499644 &$N=1-0,J=3/2-1/2,F=1/2-1/2$ &  5.4 &  1.25 & --    &              -- &              -- & -- \\
$^{12}$CO      &115.271202 &$1-0$                       &  5.5 & 0.012 & $1121 \pm 5$ & $54.1 \pm 0.2$  & $520.5 \pm 0.3$ & $45.9 \pm 0.6$ \\
H$_2$CO        &140.839502 &$2_{1\,2}-1_{1\,1}$         & 21.9 &  8.16 & $5 \pm 3$ & $0.16 \pm 0.07$ & $516 \pm 2$ & $31 \pm 6$ \\
H$_2^{13}$CO   &141.983740 &$2_{0\,2}-1_{0\,1}$         & 10.2 &  10.9 & $< 3$ & $< 0.06$        &                 &  \\
\enddata
\tablenotetext{a}{The errors are 3$\sigma$.  The upper limit to the integrated intensity is calculated as: $\int T_{\rm mb} dv < 3 \sigma \times \sqrt{\Delta V / \Delta v_{\rm res}} \Delta v_{\rm res}$, where $\Delta V$ is the assumed line width (40~km~s$^{-1}$) and $\Delta v_{\rm res}$ is the velocity resolution per channel.}
\label{tab2}
\end{deluxetable}

\clearpage
\begin{deluxetable}{lllllll}
\tablecolumns{7}
\tablewidth{0pt}
\tabletypesize{\footnotesize}
\tablecaption{Column Densities and Rotation Temperatures in M51~P1$^{\rm a}$.}
\tablehead{
\colhead{Source size$^{\rm b}$} & \multicolumn{2}{c}{No Corr.} & \multicolumn{2}{c}{10~arcsec} & \multicolumn{2}{c}{5~arcsec} \\ \hline
\colhead{Molecule} & \colhead{$N$ (cm$^{-2}$)} & \colhead{$T_{\rm ex}$ (K)} & \colhead{$N$ (cm$^{-2}$)} & \colhead{$T_{\rm ex}$ (K)} & \colhead{$N$ (cm$^{-2}$)} & \colhead{$T_{\rm ex}$ (K)} }
\startdata
HNCO & $(5.6 \pm 1.5) \times 10^{12} $ & $7.8 \pm 2.1$ 
     & $(7.4 \pm 3.3) \times 10^{13} $ & $4.9 \pm 0.8$ 
     & $(3.0 \pm 1.4) \times 10^{14} $ & $4.9 \pm 0.8$ \\
CH$_3$OH & $(2.4 \pm 0.1) \times 10^{13} $ & $9.1 \pm 1.2$ 
     & $(2.5 \pm 0.4) \times 10^{14} $ & $4.2 \pm 0.2$ 
     & $(9.8 \pm 1.4) \times 10^{14} $ & $4.2 \pm 0.2$ \\
CS & $(4.3 \pm 1.5) \times 10^{12}$ & $5.0 \pm 1.1$ 
     & $(8.0 \pm 6.8) \times 10^{13} $ & $3.3 \pm 0.4$ 
     & $(3.2 \pm 2.7) \times 10^{14} $ & $3.2 \pm 0.4$ \\
\enddata
\tablenotetext{a}{Obtained by a least square fit for multiple transitions.  The error denotes 3~$\sigma$.}
\tablenotetext{b}{Assumed source size.  \textit{No Corr.} indicates the results without a source size correction.}
\label{tab3}
\end{deluxetable}

\clearpage
\begin{deluxetable}{lllllll}
\rotate
\tablecolumns{7}
\tablewidth{0pt}
\tabletypesize{\scriptsize}
\tablecaption{Column Densities and Fractional Abundances of Identified Molecules in M51~P1.$^{\rm a}$}
\tablehead{
\colhead{Molecule} & \multicolumn{2}{c}{No Corr. ($T$=10~K)$^{\rm b}$} & \multicolumn{2}{c}{10~arcsec ($T$=5~K)$^{\rm b}$} & \multicolumn{2}{c}{5~arcsec ($T$=5~K)$^{\rm b}$} \\
\colhead{} & \colhead{$N$ (cm$^{-2}$)$^{\rm c}$} & \colhead{$X$$^{\rm c,d}$} & \colhead{(cm$^{-2}$)$^{\rm c}$} & \colhead{$X$$^{\rm c,d}$} & \colhead{(cm$^{-2}$)$^{\rm c}$} & \colhead{$X$$^{\rm c,d}$} }
\startdata
CCH              & $(1.9 \pm 0.5) \times 10^{13}$ & $(3.5 \pm 1.2) \times 10^{-9}$ 
                 & $(1.6 \pm 0.4) \times 10^{14}$ & $(5.0 \pm 1.8) \times 10^{-9}$ 
                 & $(6.3 \pm 1.8) \times 10^{14}$ & $(5.0 \pm 1.8) \times 10^{-9}$ \\
CN               & $(2.0 \pm 0.4) \times 10^{13}$ & $(3.6 \pm 1.1) \times 10^{-9}$ 
                 & $(1.1 \pm 0.2) \times 10^{14}$ & $(3.5 \pm 1.0) \times 10^{-9}$ 
                 & $(4.4 \pm 0.9) \times 10^{14}$ & $(3.5 \pm 1.0) \times 10^{-9}$ \\
HCN              & $(5.2 \pm 1.1) \times 10^{12}$ & $(9.5 \pm 2.8) \times 10^{-10}$ 
                 & $(4.2 \pm 0.9) \times 10^{13}$ & $(1.3 \pm 0.4) \times 10^{-9}$ 
                 & $(1.7 \pm 0.3) \times 10^{14}$ & $(1.3 \pm 0.4) \times 10^{-9}$ \\
H$^{13}$CN       & $(2.0 \pm 1.4) \times 10^{11}$ & $(3.6 \pm 2.7) \times 10^{-11}$ 
                 & $(1.7 \pm 1.2) \times 10^{12}$ & $(5.3 \pm 4.0) \times 10^{-11}$ 
                 & $(6.7 \pm 4.9) \times 10^{12}$ & $(5.3 \pm 4.0) \times 10^{-11}$ \\
DCN              & $< 6.4 \times 10^{10}$         & $< 1.2 \times 10^{-11}$ 
                 & $< 3.3 \times 10^{11}$         & $< 1.0 \times 10^{-11}$ 
                 & $< 1.3 \times 10^{12}$         & $< 1.0 \times 10^{-11}$ \\
HNC              & $(1.7 \pm 0.4) \times 10^{12}$ & $(3.1 \pm 0.9) \times 10^{-10}$ 
                 & $(1.3 \pm 0.3) \times 10^{13}$ & $(4.2 \pm 1.2) \times 10^{-10}$ 
                 & $(5.2 \pm 1.1) \times 10^{13}$ & $(4.2 \pm 1.2) \times 10^{-10}$ \\
HN$^{13}$C       & $< 9.4 \times 10^{10}$         & $< 1.7 \times 10^{-11}$  
                 & $< 7.8 \times 10^{11}$         & $< 2.5 \times 10^{-11}$  
                 & $< 3.1 \times 10^{12}$         & $< 2.5 \times 10^{-11}$  \\
$^{12}$CO        & $(6.4 \pm 1.3) \times 10^{16}$ & $(1.2 \pm 0.3) \times 10^{-5}$ 
                 & $(3.4 \pm 0.7) \times 10^{17}$ & $(1.1 \pm 0.3) \times 10^{-5}$ 
                 & $(1.4 \pm 0.3) \times 10^{18}$ & $(1.1 \pm 0.3) \times 10^{-5}$ \\
$^{13}$CO        & $(7.4 \pm 1.5) \times 10^{15}$ & $(1.3 \pm 0.4) \times 10^{-6}$ 
                 & $(4.2 \pm 0.8) \times 10^{16}$ & $(1.3 \pm 0.4) \times 10^{-6}$ 
                 & $(1.7 \pm 0.3) \times 10^{17}$ & $(1.3 \pm 0.4) \times 10^{-6}$ \\
C$^{17}$O        & $(1.5 \pm 0.6) \times 10^{14}$ & $(2.8 \pm 1.2) \times 10^{-8}$ 
                 & $(8.3 \pm 3.1) \times 10^{14}$ & $(2.6 \pm 1.1) \times 10^{-8}$ 
                 & $(3.3 \pm 1.2) \times 10^{15}$ & $(2.6 \pm 1.1) \times 10^{-8}$ \\
C$^{18}$O        & $(1.9 \pm 0.4) \times 10^{15}$ &  
                 & $(1.1 \pm 0.2) \times 10^{16}$ &  
                 & $(4.3 \pm 0.9) \times 10^{16}$ &  \\
HCO$^{+}$        & $(2.2 \pm 0.5) \times 10^{12}$ & $(4.1 \pm 1.2) \times 10^{-10}$ 
                 & $(1.8 \pm 0.4) \times 10^{13}$ & $(5.7 \pm 1.7) \times 10^{-10}$ 
                 & $(7.2 \pm 1.5) \times 10^{13}$ & $(5.7 \pm 1.7) \times 10^{-10}$ \\
H$^{13}$CO$^{+}$ & $(7.2 \pm 6.5) \times 10^{10}$ & $(1.3 \pm 1.2) \times 10^{-11}$ 
                 & $(6.0 \pm 5.4) \times 10^{11}$ & $(1.9 \pm 1.8) \times 10^{-11}$ 
                 & $(2.4 \pm 2.2) \times 10^{12}$ & $(1.9 \pm 1.8) \times 10^{-11}$ \\
DCO$^{+}$        & $< 3.6 \times 10^{10}$         & $< 6.5 \times 10^{-12}$ 
                 & $< 2.1 \times 10^{11}$         & $< 6.8 \times 10^{-12}$ 
                 & $< 8.6 \times 10^{11}$         & $< 6.8 \times 10^{-12}$ \\
H$_{2}$CO        & $(1.8 \pm 0.5) \times 10^{12}$ & $(3.2 \pm 1.1) \times 10^{-10}$ 
                 & $(7.8 \pm 1.8) \times 10^{12}$ & $(2.5 \pm 0.8) \times 10^{-10}$ 
                 & $(3.1 \pm 0.7) \times 10^{13}$ & $(2.5 \pm 0.8) \times 10^{-10}$ \\
H$_{2}^{13}$CO   & $< 3.3 \times 10^{11}$         & $< 6.0 \times 10^{-11}$ 
                 & $< 1.7 \times 10^{12}$         & $< 5.4 \times 10^{-11}$ 
                 & $< 6.8 \times 10^{12}$         & $< 5.4 \times 10^{-11}$ \\
HDCO             & $< 4.9 \times 10^{11}$         & $< 8.8 \times 10^{-11}$ 
                 & $< 4.2 \times 10^{12}$         & $< 1.3 \times 10^{-10}$ 
                 & $< 1.7 \times 10^{13}$         & $< 1.3 \times 10^{-10}$ \\
CH$_{3}$OH       & $(1.7 \pm 0.4) \times 10^{13}$ & $(3.1 \pm 1.0) \times 10^{-9}$ 
                 & $(1.1 \pm 0.2) \times 10^{14}$ & $(3.5 \pm 1.1) \times 10^{-9}$ 
                 & $(4.4 \pm 1.0) \times 10^{14}$ & $(3.5 \pm 1.1) \times 10^{-9}$ \\
N$_{2}$H$^{+}$   & $(7.0 \pm 1.6) \times 10^{11}$ & $(1.3 \pm 0.4) \times 10^{-10}$ 
                 & $(5.7 \pm 1.3) \times 10^{12}$ & $(1.8 \pm 0.6) \times 10^{-10}$ 
                 & $(2.3 \pm 0.5) \times 10^{13}$ & $(1.8 \pm 0.6) \times 10^{-10}$ \\
SiO              & $< 8.7 \times 10^{11}$         & $< 1.6 \times 10^{-11}$ 
                 & $< 8.6 \times 10^{11}$         & $< 2.7 \times 10^{-11}$ 
                 & $< 3.4 \times 10^{12}$         & $< 2.7 \times 10^{-11}$ \\
HNCO             & $(5.0 \pm 1.3) \times 10^{12}$ & $(9.1 \pm 3.0) \times 10^{-10}$ 
                 & $(7.2 \pm 1.8) \times 10^{13}$ & $(2.3 \pm 0.7) \times 10^{-9}$ 
                 & $(2.9 \pm 0.7) \times 10^{14}$ & $(2.3 \pm 0.7) \times 10^{-9}$ \\
c-C$_{3}$H$_{2}$ & $(5.8 \pm 4.5) \times 10^{11}$ & $(1.1 \pm 0.8) \times 10^{-10}$ 
                 & $(4.2 \pm 3.3) \times 10^{12}$ & $(1.3 \pm 1.1) \times 10^{-10}$ 
                 & $(1.7 \pm 1.3) \times 10^{13}$ & $(1.3 \pm 1.1) \times 10^{-10}$ \\
CH$_{3}$CCH      & $< 2.5 \times 10^{12}$         & $< 4.6 \times 10^{-10}$  
                 & $< 4.6 \times 10^{13}$         & $< 1.5 \times 10^{-9}$  
                 & $< 1.9 \times 10^{14}$         & $< 1.5 \times 10^{-9}$  \\
CH$_{3}$CN       & $< 1.2 \times 10^{11}$         & $< 2.2 \times 10^{-11}$  
                 & $< 1.6 \times 10^{12}$         & $< 5.2 \times 10^{-11}$  
                 & $< 6.5 \times 10^{12}$         & $< 5.2 \times 10^{-11}$  \\
CS               & $(3.1 \pm 0.7) \times 10^{12}$ & $(5.7 \pm 1.8) \times 10^{-10}$ 
                 & $(2.6 \pm 0.6) \times 10^{13}$ & $(8.2 \pm 2.6) \times 10^{-10}$ 
                 & $(1.0 \pm 0.2) \times 10^{14}$ & $(8.2 \pm 2.6) \times 10^{-10}$ \\
C$^{34}$S        & $(2.8 \pm 1.5) \times 10^{11}$ & $(5.1 \pm 3.0) \times 10^{-11}$ 
                 & $(2.4 \pm 1.3) \times 10^{12}$ & $(7.6 \pm 4.4) \times 10^{-11}$ 
                 & $(9.6 \pm 5.2) \times 10^{12}$ & $(7.6 \pm 4.4) \times 10^{-11}$ \\
H$_{2}$CS        & $< 1.1 \times 10^{12}$         & $< 2.0 \times 10^{-10}$  
                 & $< 1.1 \times 10^{13}$         & $< 3.4 \times 10^{-10}$  
                 & $< 4.3 \times 10^{13}$         & $< 3.4 \times 10^{-10}$  \\
SO               & $(1.1 \pm 0.6) \times 10^{12}$ & $(2.0 \pm 1.2) \times 10^{-10}$ 
                 & $(9.1 \pm 5.3) \times 10^{12}$ & $(2.9 \pm 1.8) \times 10^{-10}$ 
                 & $(3.6 \pm 2.1) \times 10^{13}$ & $(2.9 \pm 1.8) \times 10^{-10}$ \\
HC$_{3}$N        & $< 3.1 \times 10^{11}$         & $< 5.7 \times 10^{-11}$  
                 & $< 1.6 \times 10^{13}$         & $< 5.1 \times 10^{-10}$  
                 & $< 6.4 \times 10^{13}$         & $< 5.1 \times 10^{-10}$  \\
OCS              & $< 5.7 \times 10^{12}$         & $< 1.0 \times 10^{-9}$  
                 & $< 1.6 \times 10^{14}$         & $< 5.0 \times 10^{-9}$  
                 & $< 6.3 \times 10^{14}$         & $< 5.0 \times 10^{-9}$  \\
SO$_{2}$         & $< 8.1 \times 10^{12}$         & $< 1.5 \times 10^{-10}$  
                 & $< 4.4 \times 10^{12}$         & $< 1.4 \times 10^{-10}$  
                 & $< 1.7 \times 10^{13}$         & $< 1.4 \times 10^{-10}$  \\
\enddata
\tablenotetext{a}{The errors of column densities are estimated by taking into account the r.m.s. noise and calibration uncertainties of the chopper-wheel method (20~\%).}
\tablenotetext{b}{Assumed source sizes and excitation temperatures.  `No Corr.' indicates that the column densities are estimated without taking into account of the source size.}
\tablenotetext{c}{The upper limit to the column density is estimated from the 3$\sigma$ upper limit of the integrated intensity assuming the line width of 40~km/s both for M51~P1 and P2.}
\tablenotetext{d}{The H$_{2}$ column densities are estimated from C$^{18}$O, where the $N({\rm H_{2}})/N({\rm C^{18}O})$ ratio is assumed to be $2.9 \times 10^{6}$ \citep{meier05}.}
\label{tab4}
\end{deluxetable}

\clearpage
\begin{deluxetable}{lllllll}
\rotate
\tablecolumns{7}
\tablewidth{0pt}
\tabletypesize{\scriptsize}
\tablecaption{Column Densities and Fractional Abundances of Identified Molecules in M51~P2.$^{\rm a}$}
\tablehead{
\colhead{Molecule} & \multicolumn{2}{c}{No Corr. ($T$=10~K)$^{\rm b}$} & \multicolumn{2}{c}{10~arcsec ($T$=5~K)$^{\rm b}$} & \multicolumn{2}{c}{5~arcsec ($T$=5~K)$^{\rm b}$} \\
\colhead{} & \colhead{$N$ (cm$^{-2}$)$^{\rm c}$} & \colhead{$X$$^{\rm c,d}$} & \colhead{(cm$^{-2}$)$^{\rm c}$} & \colhead{$X$$^{\rm c,d}$} & \colhead{(cm$^{-2}$)$^{\rm c}$} & \colhead{$X$$^{\rm c,d}$} }
\startdata
CCH              & $(1.0 \pm 0.4) \times 10^{13}$ & $(3.1 \pm 1.5) \times 10^{-9}$
                 & $(8.7 \pm 3.6) \times 10^{13}$ & $(4.5 \pm 2.1) \times 10^{-9}$ 
                 & $(3.5 \pm 1.4) \times 10^{14}$ & $(4.5 \pm 2.1) \times 10^{-9}$\\
CN               & $(7.4 \pm 2.0) \times 10^{12}$ & $(2.2 \pm 0.7) \times 10^{-9}$
                 & $(4.5 \pm 1.6) \times 10^{13}$ & $(2.3 \pm 1.0) \times 10^{-9}$ 
                 & $(1.8 \pm 0.7) \times 10^{14}$ & $(2.3 \pm 1.0) \times 10^{-9}$\\
HCN              & $(3.2 \pm 0.7) \times 10^{12}$ & $(9.6 \pm 2.9) \times 10^{-10}$
                 & $(2.6 \pm 0.5) \times 10^{13}$ & $(1.4 \pm 0.4) \times 10^{-9}$ 
                 & $(1.0 \pm 0.2) \times 10^{14}$ & $(1.4 \pm 0.4) \times 10^{-9}$\\
H$^{13}$CN       & $< 1.4 \times 10^{11}$         & $< 4.1 \times 10^{-11}$ 
                 & $< 1.2 \times 10^{12}$         & $< 6.0 \times 10^{-11}$ 
                 & $< 4.6 \times 10^{12}$         & $< 6.0 \times 10^{-11}$ \\
HNC              & $(1.1 \pm 0.3) \times 10^{12}$ & $(3.3 \pm 1.0) \times 10^{-10}$
                 & $(8.6 \pm 2.0) \times 10^{12}$ & $(4.4 \pm 1.4) \times 10^{-10}$ 
                 & $(3.4 \pm 0.8) \times 10^{13}$ & $(4.4 \pm 1.4) \times 10^{-10}$\\
HN$^{13}$C       & $< 1.4 \times 10^{11}$         & $< 4.3 \times 10^{-11}$ 
                 & $< 1.2 \times 10^{12}$         & $< 6.2 \times 10^{-11}$  
                 & $< 4.8 \times 10^{12}$         & $< 6.2 \times 10^{-11}$ \\
$^{12}$CO        & $(5.0 \pm 1.0) \times 10^{16}$ & $(1.5 \pm 0.4) \times 10^{-5}$
                 & $(2.7 \pm 0.5) \times 10^{17}$ & $(1.4 \pm 0.4) \times 10^{-5}$ 
                 & $(1.1 \pm 0.2) \times 10^{18}$ & $(1.4 \pm 0.4) \times 10^{-5}$\\
$^{13}$CO        & $(5.1 \pm 1.0) \times 10^{15}$ & $(1.5 \pm 0.4) \times 10^{-6}$
                 & $(2.9 \pm 0.6) \times 10^{16}$ & $(1.5 \pm 0.4) \times 10^{-6}$ 
                 & $(1.2 \pm 0.2) \times 10^{17}$ & $(1.5 \pm 0.4) \times 10^{-6}$\\
C$^{17}$O        & $(1.2 \pm 0.8) \times 10^{14}$ & $(3.6 \pm 2.5) \times 10^{-8}$
                 & $(6.6 \pm 4.4) \times 10^{14}$ & $(3.4 \pm 2.4) \times 10^{-8}$ 
                 & $(2.6 \pm 1.7) \times 10^{15}$ & $(3.4 \pm 2.4) \times 10^{-8}$\\
C$^{18}$O        & $(1.2 \pm 0.2) \times 10^{15}$ &  
                 & $(6.6 \pm 1.4) \times 10^{15}$ &  
                 & $(2.6 \pm 0.6) \times 10^{16}$ & \\
HCO$^{+}$        & $(1.5 \pm 0.3) \times 10^{12}$ & $(4.5 \pm 1.4) \times 10^{-10}$
                 & $(1.2 \pm 0.3) \times 10^{13}$ & $(6.3 \pm 1.9) \times 10^{-10}$ 
                 & $(4.9 \pm 1.0) \times 10^{13}$ & $(6.3 \pm 1.9) \times 10^{-10}$\\
H$^{13}$CO$^{+}$ & $< 9.3 \times 10^{10}$         & $< 2.8 \times 10^{-11}$ 
                 & $< 7.8 \times 10^{11}$         & $< 4.0 \times 10^{-11}$ 
                 & $< 3.1 \times 10^{12}$         & $< 4.0 \times 10^{-11}$ \\
H$_{2}$CO        & $(9.8 \pm 4.7) \times 10^{11}$ & $(2.9 \pm 1.5) \times 10^{-10}$
                 & $(4.2 \pm 2.0) \times 10^{12}$ & $(2.2 \pm 1.2) \times 10^{-10}$ 
                 & $(1.7 \pm 0.8) \times 10^{13}$ & $(2.2 \pm 1.2) \times 10^{-10}$\\
H$_{2}^{13}$CO   & $< 5.5 \times 10^{11}$         & $< 1.6 \times 10^{-10}$ 
                 & $< 2.9 \times 10^{12}$         & $< 1.5 \times 10^{-10}$ 
                 & $< 1.1 \times 10^{13}$         & $< 1.5 \times 10^{-10}$ \\
CH$_{3}$OH       & $(1.0 \pm 0.3) \times 10^{13}$ & $(3.1 \pm 1.2) \times 10^{-9}$
                 & $(6.7 \pm 2.2) \times 10^{13}$ & $(3.5 \pm 1.4) \times 10^{-9}$ 
                 & $(2.7 \pm 0.9) \times 10^{14}$ & $(3.5 \pm 1.4) \times 10^{-9}$\\
N$_{2}$H$^{+}$   & $(3.2 \pm 1.2) \times 10^{11}$ & $(9.5 \pm 4.1) \times 10^{-11}$
                 & $(2.6 \pm 1.0) \times 10^{12}$ & $(1.4 \pm 0.6) \times 10^{-10}$ 
                 & $(1.1 \pm 0.4) \times 10^{13}$ & $(1.4 \pm 0.6) \times 10^{-10}$\\
SiO              & $< 1.4 \times 10^{11}$         & $< 4.1 \times 10^{-11}$ 
                 & $< 1.4 \times 10^{12}$         & $< 7.1 \times 10^{-11}$ 
                 & $< 5.5 \times 10^{12}$         & $< 7.1 \times 10^{-11}$ \\
HNCO             & $(2.4 \pm 1.0) \times 10^{12}$ & $(7.1 \pm 3.3) \times 10^{-10}$
                 & $(3.4 \pm 1.4) \times 10^{13}$ & $(1.8 \pm 0.8) \times 10^{-9}$ 
                 & $(1.4 \pm 0.6) \times 10^{14}$ & $(1.8 \pm 0.8) \times 10^{-9}$\\
c-C$_{3}$H$_{2}$ & $< 5.8 \times 10^{11}$         & $< 1.7 \times 10^{-10}$ 
                 & $< 4.2 \times 10^{12}$         & $< 2.2 \times 10^{-10}$ 
                 & $< 1.7 \times 10^{13}$         & $< 2.2 \times 10^{-10}$ \\
CH$_{3}$CCH      & $< 4.3 \times 10^{12}$         & $< 1.3 \times 10^{-9}$ 
                 & $< 7.9 \times 10^{13}$         & $< 4.1 \times 10^{-9}$  
                 & $< 3.2 \times 10^{14}$         & $< 4.1 \times 10^{-9}$ \\
CH$_{3}$CN       & $< 1.8 \times 10^{11}$         & $< 5.2 \times 10^{-11}$ 
                 & $< 3.0 \times 10^{12}$         & $< 1.6 \times 10^{-10}$  
                 & $< 1.2 \times 10^{13}$         & $< 1.6 \times 10^{-10}$ \\
CS               & $(1.8 \pm 0.6) \times 10^{12}$ & $(5.3 \pm 2.0) \times 10^{-10}$
                 & $(1.5 \pm 0.5) \times 10^{13}$ & $(7.7 \pm 2.9) \times 10^{-10}$ 
                 & $(5.9 \pm 1.8) \times 10^{13}$ & $(7.7 \pm 2.9) \times 10^{-10}$\\
C$^{34}$S        & $< 3.5 \times 10^{11}$         & $< 1.0 \times 10^{-10}$ 
                 & $< 3.0 \times 10^{12}$         & $< 1.5 \times 10^{-10}$ 
                 & $< 1.2 \times 10^{13}$         & $< 1.5 \times 10^{-10}$ \\
H$_{2}$CS        & $< 4.0 \times 10^{12}$         & $< 1.2 \times 10^{-9}$ 
                 & $< 3.9 \times 10^{13}$         & $< 2.0 \times 10^{-9}$  
                 & $< 1.6 \times 10^{14}$         & $< 2.0 \times 10^{-9}$ \\
SO               & $< 4.6 \times 10^{11}$         & $< 1.4 \times 10^{-10}$ 
                 & $< 2.4 \times 10^{12}$         & $< 1.2 \times 10^{-10}$ 
                 & $< 9.5 \times 10^{12}$         & $< 1.2 \times 10^{-10}$ \\
HC$_{3}$N        & $< 5.1 \times 10^{11}$         & $< 1.5 \times 10^{-10}$ 
                 & $< 2.6 \times 10^{13}$         & $< 1.4 \times 10^{-9}$  
                 & $< 1.0 \times 10^{14}$         & $< 1.4 \times 10^{-9}$ \\
OCS              & $< 3.6 \times 10^{12}$         & $< 1.1 \times 10^{-9}$ 
                 & $< 1.0 \times 10^{14}$         & $< 5.2 \times 10^{-9}$  
                 & $< 4.0 \times 10^{14}$         & $< 5.2 \times 10^{-9}$ \\
SO$_{2}$         & $< 3.4 \times 10^{12}$         & $< 1.0 \times 10^{-9}$ 
                 & $< 1.9 \times 10^{13}$         & $< 9.6 \times 10^{-10}$  
                 & $< 7.4 \times 10^{13}$         & $< 9.6 \times 10^{-10}$ \\
\enddata
\tablenotetext{a}{The errors of column densities are estimated by taking into account the r.m.s. noise and calibration uncertainties of the chopper-wheel method (20~\%).}
\tablenotetext{b}{Assumed source sizes and excitation temperatures.  `No Corr.' indicates that the column densities are estimated without taking into account of the source size.}
\tablenotetext{c}{The upper limit to the column density is estimated from the 3$\sigma$ upper limit of the integrated intensity assuming the line width of 40~km/s both for M51~P1 and P2.}
\tablenotetext{d}{The H$_{2}$ column densities are estimated from C$^{18}$O, where the $N({\rm H_{2}})/N({\rm C^{18}O})$ ratio is assumed to be $2.9 \times 10^{6}$ \citep{meier05}.}
\label{tab4b}
\end{deluxetable}

\clearpage
\begin{deluxetable}{lll}
\tablecolumns{3}
\tablewidth{0pt}
\tablecaption{Ratios of CO Isotopologues}
\tablehead{
\colhead{Molecule} & \colhead{M51~P1} & \colhead{M51~P2}
}
\startdata
$^{13}$CO/C$^{18}$O & $4 \pm 1 $ & $4 \pm 1$ \\
C$^{18}$O/C$^{17}$O & $13 \pm 6$ & $10 \pm 7$ \\
\enddata
\label{tab5}
\end{deluxetable}

\begin{deluxetable}{lll}
\tablecolumns{3}
\tablewidth{0pt}
\tablecaption{SFR and SFE in P1 and P2 $^{\rm a}$}
\tablehead{
\colhead{ } & \colhead{M51~P1} & \colhead{M51~P2}
}
\startdata
SFR & $(0.055 \pm 0.008)$ M$_{\odot}$~yr$^{-1}$ & $(0.022 \pm 0.004)$ M$_{\odot}$~yr$^{-1}$ \\
Molecular Gas Mass & $(9.4 \pm 2.0) \times 10^{7}$~M$_{\odot}$ & $(5.7 \pm 1.2) \times 10^{7}$~M$_{\odot}$ \\
SFE & $(5.9 \pm 1.5) \times 10^{-10}$~yr$^{-1}$ & $(3.9 \pm 1.0) \times 10^{-10}$~yr$^{-1}$ \\
\enddata
\tablenotetext{a}{Derived by the method described in the Appendix~B.}
\label{tab6}
\end{deluxetable}

\clearpage
\appendix
\section{Expanded Spectra}

\begin{figure}
\epsscale{1.00}
\plotone{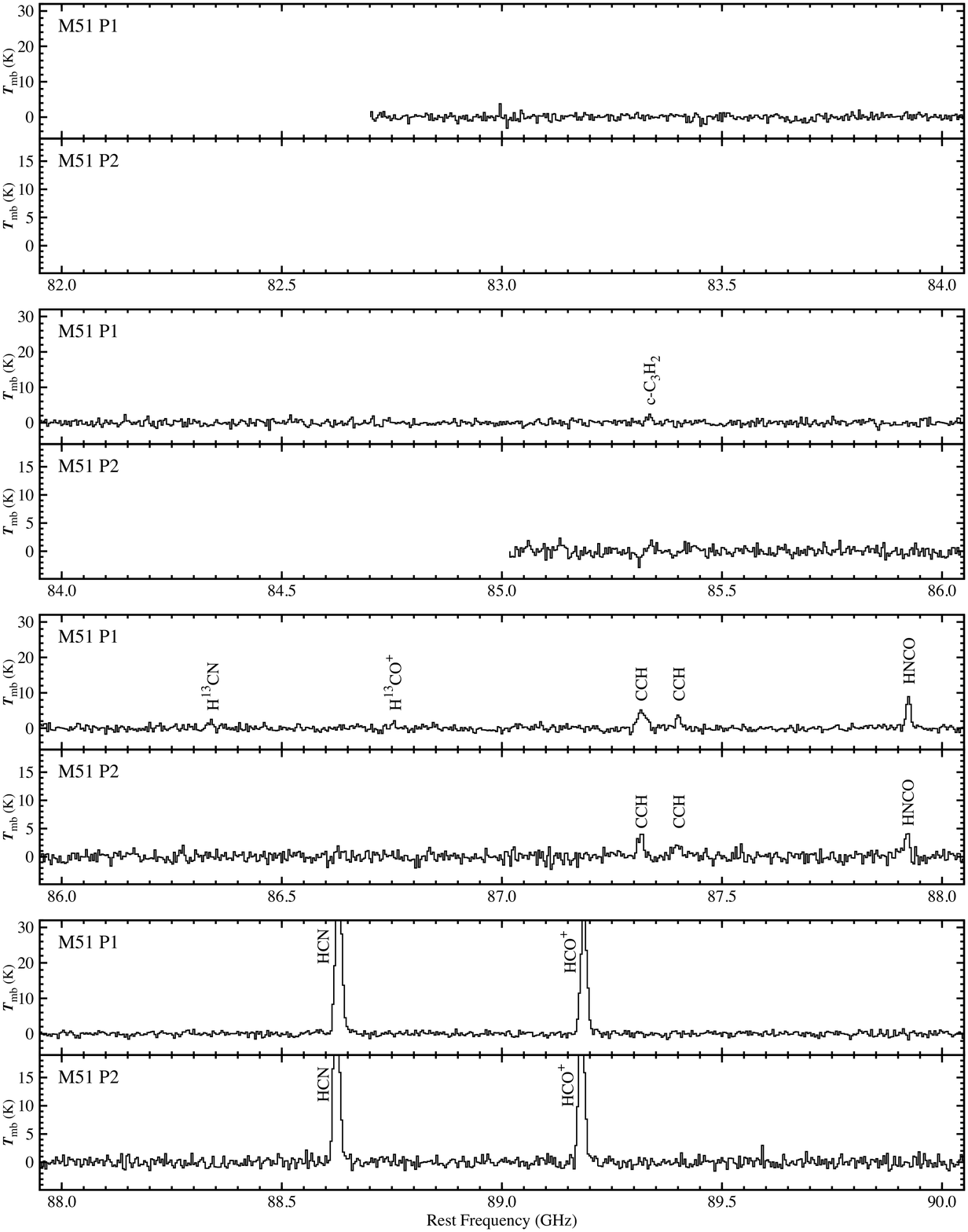}
\caption{Spectra of M51~P1 and P2.  The frequency resolution is 4~MHz.}
\label{fig2}
\end{figure}
\setcounter{figure}{6}

\clearpage
\begin{figure}
\epsscale{1.00}
\plotone{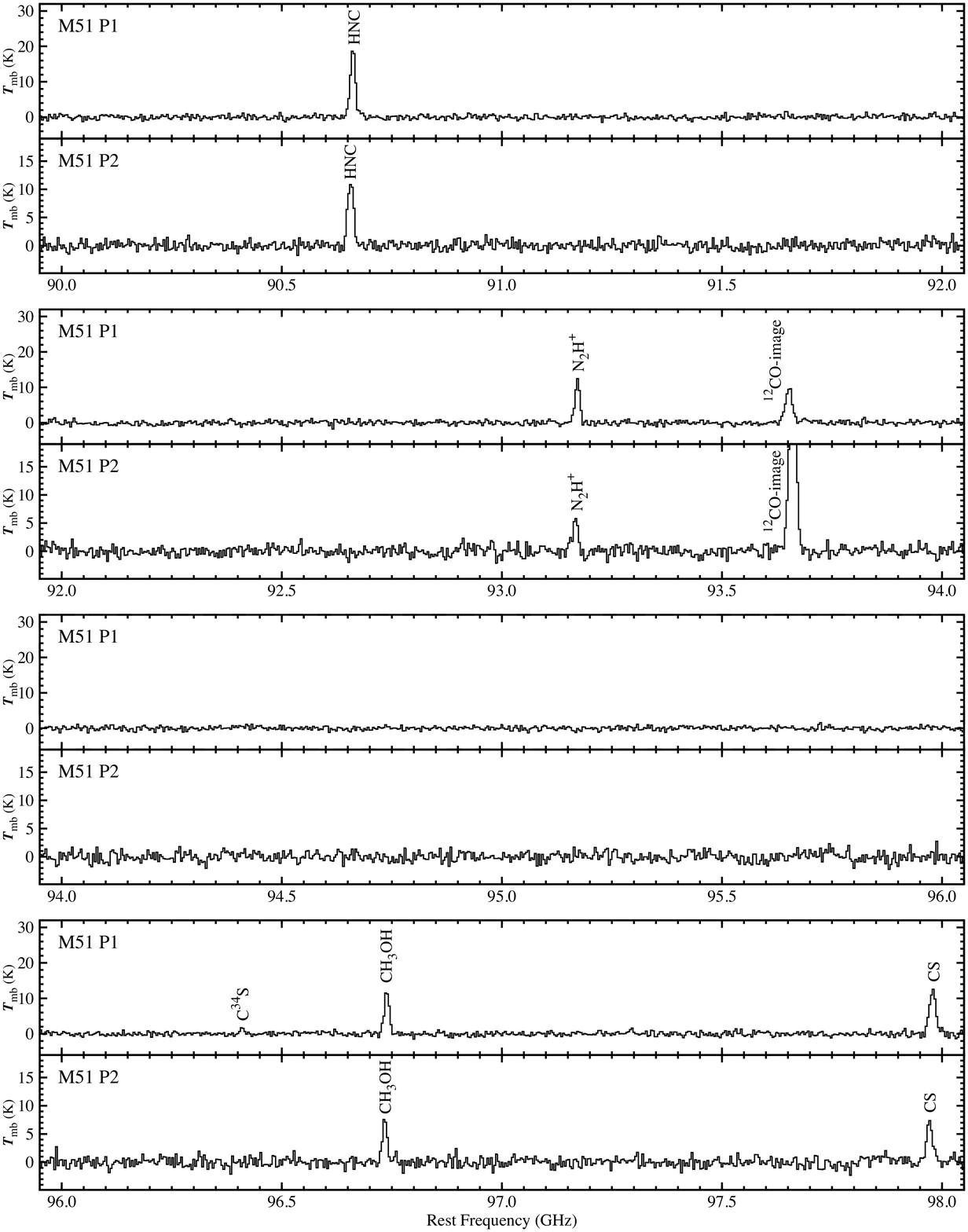}
\caption{\textit{Continued}}
\end{figure}
\setcounter{figure}{6}

\clearpage
\begin{figure}
\epsscale{1.00}
\plotone{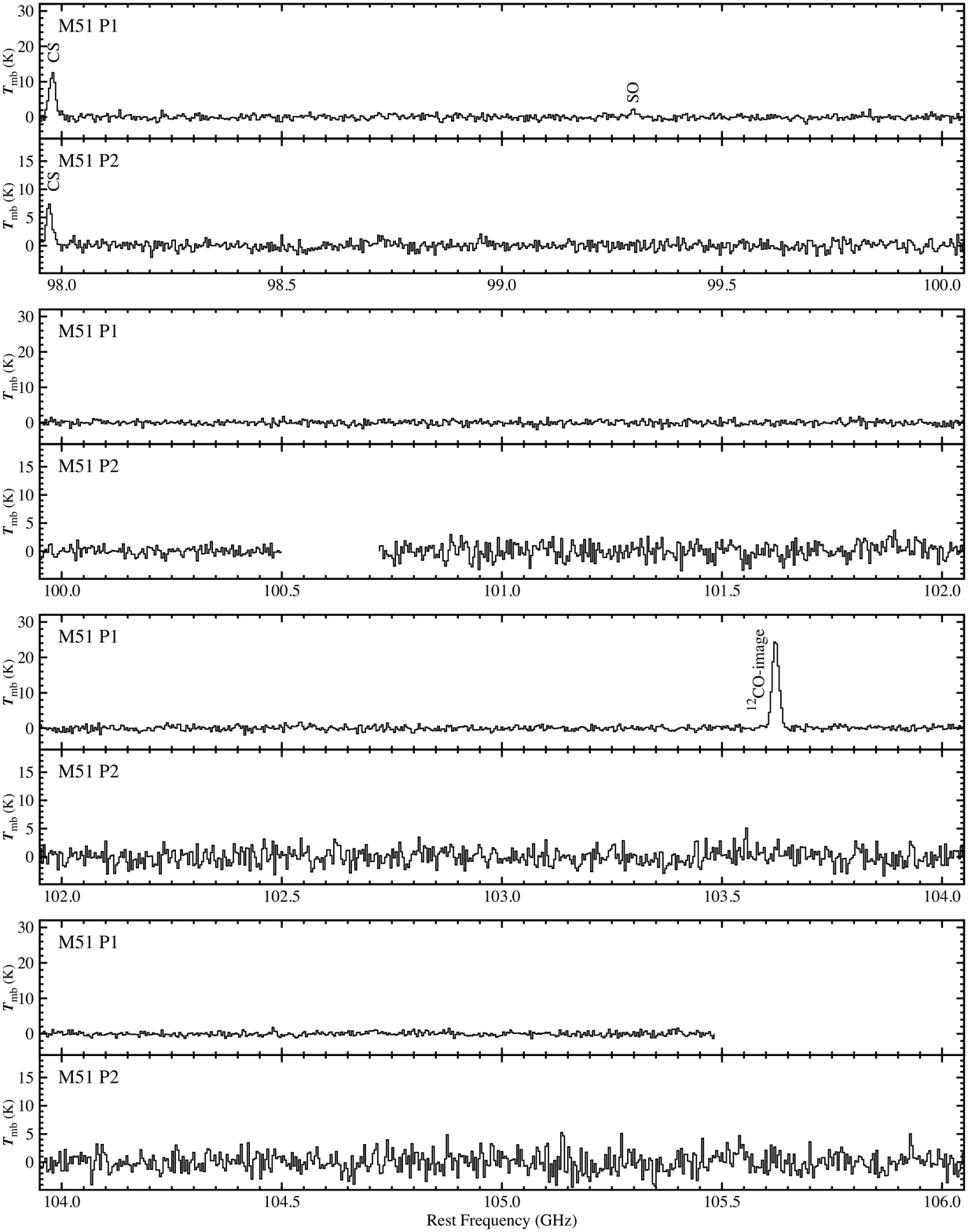}
\caption{\textit{Continued}}
\end{figure}
\setcounter{figure}{6}

\clearpage
\begin{figure}
\epsscale{1.00}
\plotone{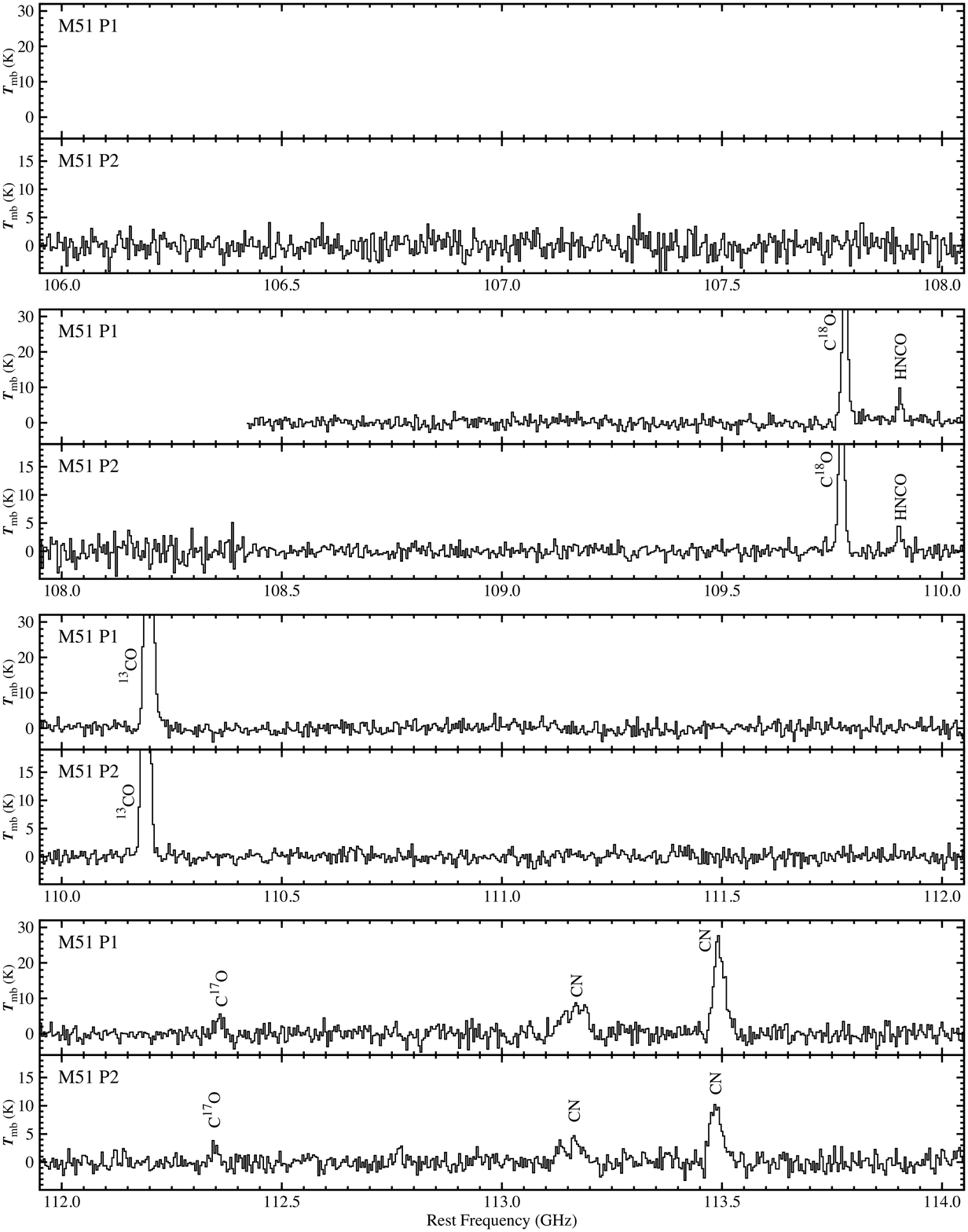}
\caption{\textit{Continued}}
\end{figure}
\setcounter{figure}{6}

\clearpage
\begin{figure}
\epsscale{1.00}
\plotone{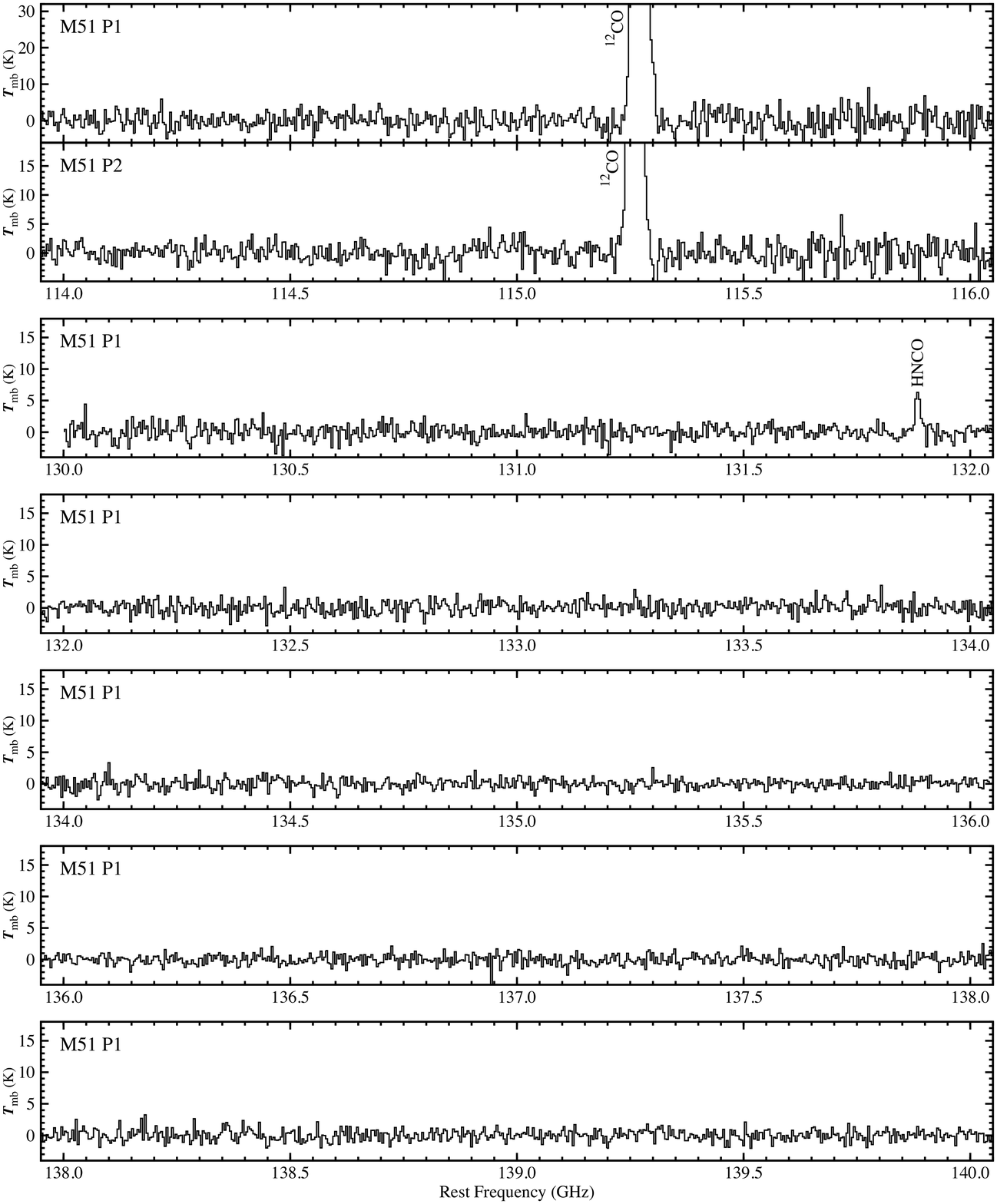}
\caption{\textit{Continued}}
\end{figure}
\setcounter{figure}{6}

\clearpage
\begin{figure}
\epsscale{1.00}
\plotone{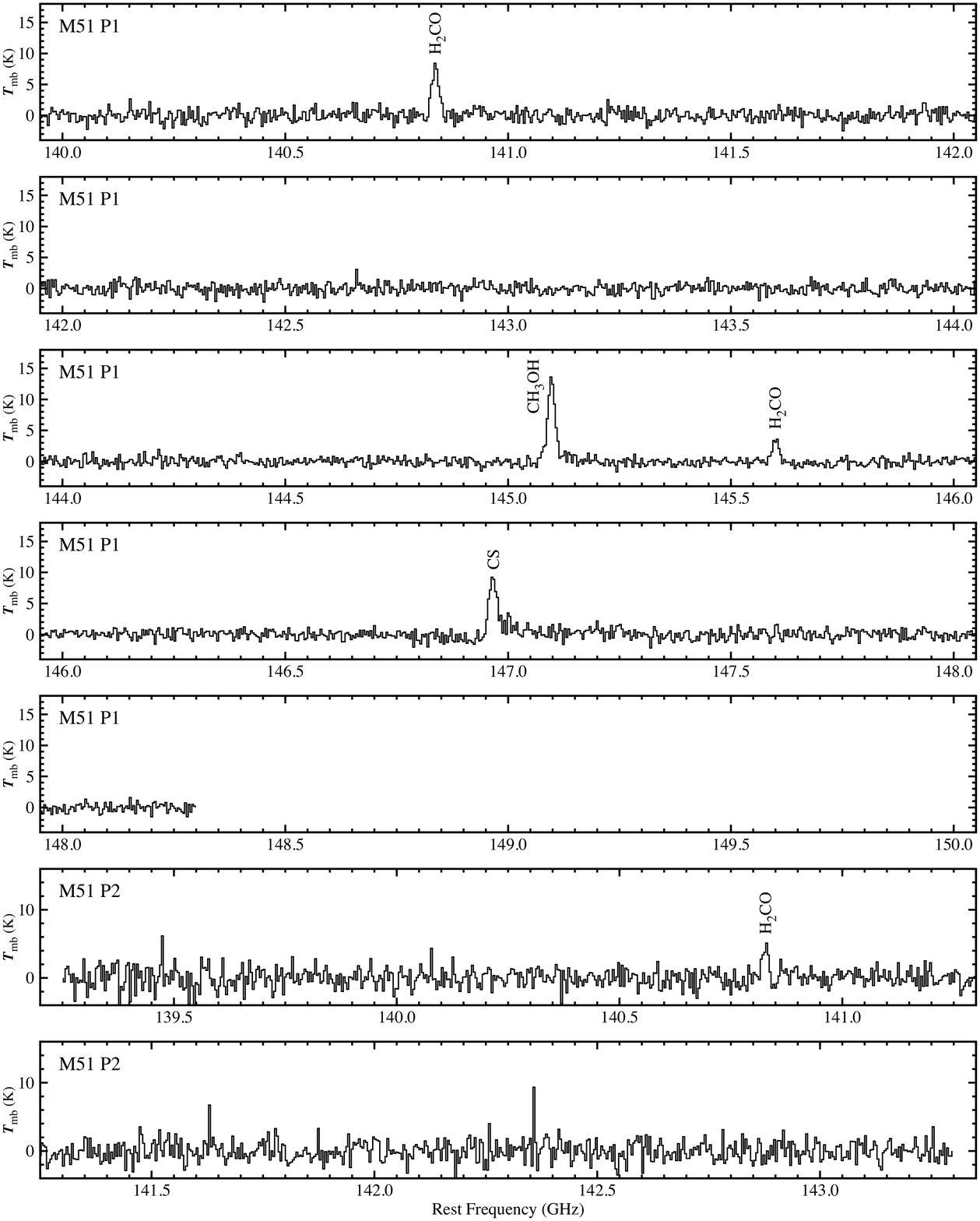}
\caption{\textit{Continued}}
\end{figure}

\clearpage
\section{Star Formation Rate and Star Formation Efficiency}
The star formation rate (SFR) was estimated from the luminosities of the H$\alpha$ emission and the 24~$\mu$m emission.  We used the H$\alpha$ data and the 24~$\mu$m data, which are provided by SINGS \citep{kennicutt2003}.  The H$\alpha$ and 24~$\mu$m data were observed with the 2.1~m telescope in Kit Peak National Observatory and the Multiband Imaging Photometer for Spitzer (MIPS), respectively.  The luminosities were estimated by measuring fluxes within the 12.5~arcsec radius in P1 and P2 where the distance is assumed to be 8.2~Mpc \citep{feldmeier1997, vink12}.  In the flux measurement, we do not apply a correction for [O$_{\rm III}$] contribution to the H$\alpha$ flux.  The SFRs were estimated by using the calibration method given by \citet{calzetti2007}:
\begin{equation}
{\rm SFR}(M_{\odot}\,{\rm yr}^{-1}) = 5.3 \times 10^{-42}[L({\rm H\alpha})_{\rm obs} + (0.031 \pm 0.006)\times L(24\,\mu m)],
\label{eqa1}
\end{equation}
where $L({\rm H\alpha})_{\rm obs}$ and $L(24\,\mu m)$ are observed luminosities of {\bf H${\alpha}$} and 24~$\mu$m, respectively.  The SFRs are estimated to be $0.055 \pm 0.008$~M$_{\odot}$yr$^{-1}$ and $0.022 \pm 0.004$~M$_{\odot}$yr$^{-1}$ for P1 and P2, respectively.

The star formation efficiency (SFE) was obtained by dividing the SFR by molecular gas mass.  The molecular gas mass was estimated from the column density of H$_2$ derived from the C$^{18}$O data taken with {\bf a} beam size of 25 arcsec and {\bf assuming a} rotation temperature of 10~K.  In this analysis, a contribution of He to the molecular gas mass was taken into account by applying {\bf a} factor of 1.37.  The molecular gas mass were estimated to be $(9.4 \pm 2.0) \times 10^{7}$~M$_{\odot}$ and $(5.7 \pm 1.2) \times 10^{7}$~M$_{\odot}$ in P1 and P2, respectively.  Then, the SFEs are estimated to be $(5.9 \pm 1.5) \times 10^{-10}$~M$_{\odot}$yr$^{-1}$ and $(3.9 \pm 1.0) \times 10^{-10}$~M$_{\odot}$yr$^{-1}$ for P1 and P2, respectively.   


\begin{thebibliography}{}
\bibitem[Aikawa(2013)]{Aikawa2013} Aikawa, Y.\ 2013, Chemical Reviews, 113, 8961 
\bibitem[Aalto et al.(1999)]{aalto1999} Aalto, S., H{\"u}ttemeister, S., Scoville, N.~Z., \& Thaddeus, P.\ 1999, \apj, 522, 165
\bibitem[Aalto et al.(2002)]{aalto2002} Aalto, S., Polatidis, A.~G., H{\"u}ttemeister, S., \& Curran, S.~J.\ 2002, \aap, 381, 783
\bibitem[Aladro et al.(2011a)]{aladro2011a} Aladro, R., Mart{\'{\i}}n-Pintado, J., Mart{\'{\i}}n, S., Mauersberger, R., \& Bayet, E.\ 2011a, \aap, 525, A89
\bibitem[Aladro et al.(2011b)]{aladro2011b} Aladro, R., Mart{\'{\i}}n, S., Mart{\'{\i}}n-Pintado, J., et al.\ 2011b, \aap, 535, A84 
\bibitem[Aladro et al.(2013)]{aladro2013} Aladro, R., Viti, S., Bayet, E., et al.\ 2013, \aap, 549, A39 
\bibitem[Bachiller \& Perez Gutierrez(1997)]{bachiller1997} Bachiller, R., \& Perez Gutierrez, M.\ 1997, \apjl, 487, L93
\bibitem[Bayet et al.(2009)]{Bayet2009} Bayet, E., Aladro, R., Mart{\'{\i}}n, S., Viti, S., \& Mart{\'{\i}}n-Pintado, J.\ 2009, \apj, 707, 126 
\bibitem[Beuther et al.(2008)]{Beuther2008} Beuther, H., Semenov, D., Henning, T., \& Linz, H.\ 2008, \apjl, 675, L33 
\bibitem[Blake et al.(1987)]{Blake1987} Blake, G.~A., Sutton, E.~C., Masson, C.~R., \& Phillips, T.~G.\ 1987, \apj, 315, 621 
\bibitem[Bland-Hawthorn et al.(1997)]{Bland-Hawthorn1997} Bland-Hawthorn, J., Gallimore, J.~F., Tacconi, L.~J., et al.\ 1997, \apss, 248, 9
\bibitem[Calzetti et al.(2007)]{calzetti2007} Calzetti, D., Kennicutt, R.~C., Engelbracht, C.~W., et al.\ 2007, \apj, 666, 870 
\bibitem[Caselli et al.(1999)]{caselli1999} Caselli, P., Walmsley, C.~M., Tafalla, M., Dore, L., \& Myers, P.~C.\ 1999, \apjl, 523, L165 
\bibitem[Caselli et al.(2002)]{Caselli2002} Caselli, P., Stantcheva, T., Shalabiea, O., Shematovich, V.~I., \& Herbst, E.\ 2002, \planss, 50, 1257
\bibitem[Caselli et al.(2003)]{Caselli2003} Caselli, P., van der Tak, F.~F.~S., Ceccarelli, C., \& Bacmann, A.\ 2003, \aap, 403, L37 
\bibitem[Costagliola et al.(2011)]{costagliola2011} Costagliola, F., Aalto, S., Rodriguez, M.~I., et al.\ 2011, \aap, 528, A30 
\bibitem[Curran et al.(2001)]{curran2001} Curran, S.~J., Johansson, L.~E.~B., Bergman, P., Heikkil{\"a}, A., \& Aalto, S.\ 2001, \aap, 367, 457 
\bibitem[Dalcanton et al.(2009)]{Dalcanton2009} Dalcanton, J.~J., Williams, B.~F., Seth, A.~C., et al.\ 2009, \apjs, 183, 67 
\bibitem[Dickens et al.(2000)]{Dickens2000} Dickens, J.~E., Irvine, W.~M., Snell, R.~L., et al.\ 2000, \apj, 542, 870 
\bibitem[Drdla et al.(1989)]{Drdla1989} Drdla, K., Knapp, G.~R., \& van Dishoeck, E.~F.\ 1989, \apj, 345, 815 
\bibitem[Egusa et al.(2011)]{egusa11} Egusa, F., Koda, J., \& Scoville, N.\ 2011, \apj, 726, 85 
\bibitem[Feldmeier et al.(1997)]{feldmeier1997} Feldmeier, J.~J., Ciardullo, R., \& Jacoby, G.~H.\ 1997, \apj, 479, 231 
\bibitem[Fuente et al.(2005)]{Fuente2005} Fuente, A., Garc{\'{\i}}a-Burillo, S., Gerin, M., et al.\ 2005, \apjl, 619, L155 
\bibitem[Fuente et al.(2008)]{Fuente2008} Fuente, A., Garc{\'{\i}}a-Burillo, S., Usero, A., et al.\ 2008, \aap, 492, 675 
\bibitem[Garc{\'{\i}}a-Burillo et al.(2010)]{garcia2010} Garc{\'{\i}}a-Burillo, S., Usero, A., Fuente, A., et al.\ 2010, \aap, 519, A2 
\bibitem[Hasegawa et al.(1992)]{hasegawa1992} Hasegawa, T.~I., Herbst, E., \& Leung, C.~M.\ 1992, \apjs, 82, 167
\bibitem[Helfer et al.(2003)]{helfer2003} Helfer, T.~T., Thornley, M.~D., Regan, M.~W., et al.\ 2003, \apjs, 145, 259 
\bibitem[Henkel et al.(1988)]{henkel1988} Henkel, C., Schilke, P., \& Mauersberger, R.\ 1988, \aap, 201, L23 
\bibitem[Hirota et al.(2009)]{Hirota2009} Hirota, T., Ohishi, M., \& Yamamoto, S.\ 2009, \apj, 699, 585 
\bibitem[Huettemeister et al.(1995)]{huettemeister1995} Huettemeister, S., Henkel, C., Mauersberger, R., et al.\ 1995, \aap, 295, 571 
\bibitem[Karachentsev et al.(2004)]{Karachentsev2004} Karachentsev, I.~D., Karachentseva, V.~E., Huchtmeier, W.~K., \& Makarov, D.~I.\ 2004, \aj, 127, 2031 
\bibitem[Kennicutt et al.(2003)]{kennicutt2003} Kennicutt, R.~C., Jr., Armus, L., Bendo, G., et al.\ 2003, \pasp, 115, 928 
\bibitem[Koda et al.(2009)]{koda09} Koda, J., Scoville, N., Sawada, T., et al.\ 2009, \apjl, 700, L132 
\bibitem[Kohno et al.(2001)]{Kohno2001} Kohno, K., Matsushita, S., Vila-Vilar{\'o}, B., et al.\ 2001, The Central Kiloparsec of Starbursts and AGN: The La Palma Connection, 249, 672 
\bibitem[Kohno et al.(1996)]{kohno1996} Kohno, K., Kawabe, R., Tosaki, T., \& Okumura, S.~K.\ 1996, \apjl, 461, L29 
\bibitem[Krips et al.(2008)]{Krips2008} Krips, M., Neri, R., Garc{\'{\i}}a-Burillo, S., et al.\ 2008, \apj, 677, 262 
\bibitem[Lepp \& Dalgarno(1996)]{Lepp1996} Lepp, S., \& Dalgarno, A.\ 1996, \aap, 306, L21 
\bibitem[Lindberg et al.(2011)]{Lindberg2011} Lindberg, J.~E., Aalto, S., Costagliola, F., et al.\ 2011, \aap, 527, A150 
\bibitem[Maloney et al.(1996)]{Maloney1996} Maloney, P.~R., Hollenbach, D.~J., \& Tielens, A.~G.~G.~M.\ 1996, \apj, 466, 561 
\bibitem[Mangum et al.(1991)]{Mangum1991} Mangum, J.~G., Plambeck, R.~L., \& Wootten, A.\ 1991, \apj, 369, 169 
\bibitem[Mart{\'{\i}}n et al.(2006)]{martin2006} Mart{\'{\i}}n, S., Mauersberger, R., Mart{\'{\i}}n-Pintado, J., Henkel, C., \& Garc{\'{\i}}a-Burillo, S.\ 2006, \apjs, 164, 450 
\bibitem[Mart{\'{\i}}n et al.(2009)]{martin2009} Mart{\'{\i}}n, S., Mart{\'{\i}}n-Pintado, J., \& Mauersberger, R.\ 2009, \apj, 694, 610 
\bibitem[Mart{\'{\i}}n et al.(2010)]{Martin2010} Mart{\'{\i}}n, S., Aladro, R., Mart{\'{\i}}n-Pintado, J., \& Mauersberger, R.\ 2010, \aap, 522, A62
\bibitem[Mart{\'{\i}}n et al.(2011)]{martin2011} Mart{\'{\i}}n, S., Krips, M., Mart{\'{\i}}n-Pintado, J., et al.\ 2011, \aap, 527, A36 
\bibitem[Meier \& Turner(2005)]{meier05} Meier, D.~S., \& Turner, J.~L.\ 2005, \apj, 618, 259 
\bibitem[Meier et al.(2011)]{meier2011} Meier, D.~S., Turner, J.~L., \& Schinnerer, E.\ 2011, \aj, 142, 32 
\bibitem[Meier \& Turner(2012)]{meier2012} Meier, D.~S., \& Turner, J.~L.\ 2012, \apj, 755, 104
\bibitem[Meijerink \& Spaans(2005)]{Meijerink2005} Meijerink, R., \& Spaans, M.\ 2005, \aap, 436, 397
\bibitem[Meijerink et al.(2007)]{Meijerink2007} Meijerink, R., Spaans, M., \& Israel, F.~P.\ 2007, \aap, 461, 793 
\bibitem[M{\"u}ller et al.(2001)]{muller01} M{\"u}ller, H.~S.~P., Thorwirth, S., Roth, D.~A., \& Winnewisser, G.\ 2001, \aap, 370, L49 
\bibitem[M\"{u}ller et~al.(2005)]{muller05} M\"{u}ller,~H.~S.~P., Schl\"{o}der,~F, Stutzki,~J. \&~Winnewisser,~G., 2005, J. Mol. Struct., 742, 215
\bibitem[Nakai et al.(1994)]{nakai1994} Nakai, N., Kuno, N., Handa, T., \& Sofue, Y.\ 1994, \pasj, 46, 527 
\bibitem[Nakajima et al.(2011)]{nakajima2011} Nakajima, T., Takano, S., Kohno, K., \& Inoue, H.\ 2011, \apjl, 728, L38 
\bibitem[Nguyen et al.(1992)]{nguyen1992} Nguyen, Q.-R., Jackson, J.~M., Henkel, C., Truong, B., \& Mauersberger, R.\ 1992, \apj, 399, 521 
\bibitem[Prasad \& Tarafdar(1983)]{Prasad1983} Prasad, S.~S., \& Tarafdar, S.~P.\ 1983, \apj, 267, 603
\bibitem[P{\'e}rez-Beaupuits et al.(2009)]{Perez2009} P{\'e}rez-Beaupuits, J.~P., Spaans, M., van der Tak, F.~F.~S., et al.\ 2009, \aap, 503, 459 
\bibitem[Pickett et~al.(1998)]{pickett98} Pickett,~H.~M., Poynter,~R.~L., Cohen,~E.~A., Delitsky,~M.~L. Pearson,~J~.C., \& M\"{u}ller,~H.~S.~P., 1998, Journal of Quantitative Spectroscopy and Radiative Transfer, 60, 883
\bibitem[Roberts(1969)]{Roberts1969} Roberts, W.~W.\ 1969, \apj, 158, 123 
\bibitem[Roberts et al.(2002)]{Roberts2002} Roberts, H., Fuller, G.~A., Millar, T.~J., Hatchell, J., \& Buckle, J.~V.\ 2002, \aap, 381, 1026 
\bibitem[Rodr{\'{\i}}guez-Fern{\'a}ndez et al.(2010)]{rodriguez2010} Rodr{\'{\i}}guez-Fern{\'a}ndez, N.~J., Tafalla, M., Gueth, F., \& Bachiller, R.\ 2010, \aap, 516, A98 
\bibitem[Sage et al.(1990)]{sage1990} Sage, L.~J., Shore, S.~N., \& Solomon, P.~M.\ 1990, \apj, 351, 422 
\bibitem[Sage et al.(1991)]{sage1991} Sage, L.~J., Henkel, C., \& Mauersberger, R.\ 1991, \aap, 249, 31
\bibitem[Sage \& Ziurys(1995)]{sage1995} Sage, L.~J., \& Ziurys, L.~M.\ 1995, \apj, 447, 625 
\bibitem[Sakai et al.(2012)]{Sakai2012} Sakai, N., Ceccarelli, C., Bottinelli, S., Sakai, T., \& Yamamoto, S.\ 2012, \apj, 754, 70 
\bibitem[Schilke et al.(1992)]{Schilke1992} Schilke, P., Walmsley, C.~M., Pineau Des Forets, G., et al.\ 1992, \aap, 256, 595 
\bibitem[Schinnerer et al.(2010)]{schinnere10} Schinnerer, E., Wei{\ss}, A., Aalto, S., \& Scoville, N.~Z.\ 2010, \apj, 719, 1588 
\bibitem[Schinnerer et al.(2013)]{schinnere13} Schinnerer, E., Meidt, S.~E., Pety, J., et al.\ 2013, \apj, 779, 42 
\bibitem[Schuster et al.(2007)]{schuster2007} Schuster, K.~F., Kramer, C., Hitschfeld, M., Garcia-Burillo, S., \& Mookerjea, B.\ 2007, \aap, 461, 143
\bibitem[Sorai et al.(2002)]{sorai2002} Sorai, K., Nakai, N., Kuno, N., \& Nishiyama, K.\ 2002, \pasj, 54, 179 
\bibitem[Spitzer(1978)]{Spitzer1978} Spitzer, L.\ 1978, New York Wiley-Interscience, 1978.~333 p.,  
\bibitem[Strickland et al.(2004)]{Strickland2004} Strickland, D.~K., Heckman, T.~M., Colbert, E.~J.~M., Hoopes, C.~G., \& Weaver, K.~A.\ 2004, \apjs, 151, 193 
\bibitem[Suzuki et al.(1992)]{suzuki1992} Suzuki, H., Yamamoto, S., Ohishi, M., et al.\ 1992, \apj, 392, 551 
\bibitem[Takano et al.(2013)]{takano2013} Takano, S., Takano, T., Nakai, N., Kawaguchi, K., \& Schilke, P.\ 2013, \aap, 552, A34 
\bibitem[Tercero et al.(2010)]{tercero2010} Tercero, B., Cernicharo, J., Pardo, J.~R., \& Goicoechea, J.~R.\ 2010, \aap, 517, A96 
\bibitem[Tielens \& Whittet(1997)]{tielens1997} Tielens, A.~G.~G.~M., \& Whittet, D.~C.~B.\ 1997, IAU Symposium, 178, 45
\bibitem[Turner(2001)]{Turner2001} Turner, B.~E.\ 2001, \apjs, 136, 579 
\bibitem[Usero et al.(2004)]{usero2004} Usero, A., Garc{\'{\i}}a-Burillo, S., Fuente, A., Mart{\'{\i}}n-Pintado, J., \& Rodr{\'{\i}}guez-Fern{\'a}ndez, N.~J.\ 2004, \aap, 419, 897 
\bibitem[Usero et al.(2006)]{usero2006} Usero, A., Garc{\'{\i}}a-Burillo, S., Mart{\'{\i}}n-Pintado, J., Fuente, A., \& Neri, R.\ 2006, \aap, 448, 457 
\bibitem[Vila-Vilar{\'o}(2008)]{vila2008} Vila-Vilar{\'o}, B.\ 2008, \pasj, 60, 1231 
\bibitem[Vink{\'o} et al.(2012)]{vink12} Vink{\'o}, J., Tak{\'a}ts, K., Szalai, T., et al.\ 2012, \aap, 540, A93 
\bibitem[Wang et al.(2004)]{wang2004} Wang, M., Henkel, C., Chin, Y.-N., et al.\ 2004, \aap, 422, 883 
\bibitem[Watanabe et al.(2003)]{watanabe2003} Watanabe, N., Shiraki, T., \& Kouchi, A.\ 2003, \apjl, 588, L121
\end{thebibliography}
\end{document}